\newcommand{\ubc}{\supit{b}}
\newcommand{\mcg}{\supit{a}}
\newcommand{\cita}{\supit{c}}
\newcommand{\sydu}{\supit{e}}
\newcommand{\ut}{\supit{f}}
\newcommand{\ibm}{\supit{g}}
\newcommand{\drao}{\supit{h}}
\newcommand{\dunlap}{\supit{i}}
\newcommand{\cmu}{\supit{j}}
\newcommand{\cifar}{\supit{d}}

\documentclass[]{spie}  
\usepackage[]{graphicx}
\usepackage[caption = false]{subfig}
\usepackage{color}
\usepackage{multirow}
\usepackage{amssymb}
\graphicspath{{figures/}{images/}}

\title{Canadian Hydrogen Intensity Mapping Experiment (CHIME) Pathfinder} 

\author{ Kevin Bandura\mcg, 
Graeme E. Addison\ubc, 
Mandana Amiri\ubc, 
J. Richard Bond\cita\cifar, 
Duncan Campbell-Wilson\sydu, 
Liam Connor\cita\ut, 
Jean-Fran\c{c}ois Cliche\mcg, 
Greg Davis\ubc, 
Meiling Deng\ubc, 
Nolan Denman\ut, 
Matt Dobbs\mcg, 
Mateus Fandino\ubc,
Kenneth Gibbs\ubc, 
Adam Gilbert\mcg, 
Mark Halpern\ubc, 
David Hanna\mcg, 
Adam D. Hincks\ubc, 
Gary Hinshaw\ubc, 
Carolin H\"ofer\ubc, 
Peter Klages\ut\ibm, 
Tom L. Landecker\drao, 
Kiyoshi Masui\ubc\cifar, 
Juan Mena\mcg, 
Laura B. Newburgh\dunlap, 
Ue-Li Pen\cita, 
Jeffrey B. Peterson\cmu, 
Andre Recnik\ut, 
J. Richard Shaw\cita, 
Kris Sigurdson\ubc, 
Michael Sitwell\ubc, 
Graeme Smecher\mcg, 
Rick Smegal\ubc, 
Keith Vanderlinde\ut\dunlap, and
Don Wiebe\ubc
\skiplinehalf
\supit{a}Department of Physics, McGill University, 3600 University St, Montreal, Canada
\skiplinehalf
\supit{b}Department of Physics and Astronomy, University of British Columbia, 
6224 Agricultural Rd. Vancouver, V6T 1Z1, Canada
\skiplinehalf
\supit{c}CITA, 60 St George St, Toronto, ON, M5S 3H8, Canada
\skiplinehalf
\supit{d}Canadian Institute for Advanced Research,
CIFAR Program in Cosmology and Gravity,
Toronto, ON M5G 1Z8 
\skiplinehalf
\supit{e}Sydney Institute for Astronomy, School of Physics, University 
of Sydney, NSW 2006, Australia
\skiplinehalf
\supit{f}Department of Astronomy \& Astrophysics, University of
Toronto, 50 St George St, Toronto, ON, M5S 3H4, Canada
\skiplinehalf
\supit{g}IBM Canada
\skiplinehalf
\supit{h}National Research Council Canada, Dominion Radio Astrophysical
Observatory, Box 248, Penticton BC V2A 6J9 Canada
\skiplinehalf
\supit{i}Dunlap Institute for Astronomy \& Astrophysics, University
of Toronto, 50 St George St, Toronto, ON, M5S 3H4, Canada
\skiplinehalf
\supit{j}McWilliams Center for Cosmology, Carnegie Mellon University,
Department of Physics, 5000 Forbes Ave, Pittsburgh PA 15213, USA
\skiplinehalf
}

\authorinfo{Send correspondence to K.Bandura: E-mail: kevin.bandura@mcgill.ca \\  }

\begin{document} 
  \maketitle 

\begin{abstract}

A pathfinder version of CHIME (the Canadian Hydrogen Intensity Mapping
Experiment) is currently being commissioned at the Dominion Radio Astrophysical
Observatory (DRAO) in Penticton, BC.   The instrument is a hybrid cylindrical
interferometer designed to measure the large scale neutral hydrogen power
spectrum across the redshift range 0.8 to 2.5. The power spectrum will be used
to measure the baryon acoustic oscillation (BAO) scale across this 
poorly probed redshift range where dark energy becomes a significant contributor to the evolution of
the Universe. The instrument revives the cylinder design in radio astronomy with
a wide field survey as a primary goal. Modern low-noise amplifiers and digital
processing remove the necessity for the analog beamforming that characterized
previous designs. The Pathfinder consists of two cylinders 37\,m long by 20\,m
wide oriented north-south for a total collecting area of 1,500 square meters.
The cylinders are stationary with no moving parts, and form a transit instrument
with an instantaneous field of view of $\sim$100\,degrees by 1-2\,degrees. Each
CHIME Pathfinder cylinder has a feedline with 64 dual polarization feeds placed every
$\sim$30\,cm which Nyquist sample the north-south sky over much of the frequency band. The
signals from each dual-polarization feed are independently amplified,
filtered to 400-800\,MHz, and directly sampled at 800\,MSps using 8 bits. The
correlator is an FX design, where the Fourier transform channelization is
performed in FPGAs, which are interfaced to a set of GPUs that compute the
correlation matrix. The CHIME Pathfinder is a 1/10th scale prototype version of
CHIME and is designed to detect the BAO feature and constrain the 
distance-redshift relation.

The lessons
learned from its implementation will be used to inform and improve the final
CHIME design.

\end{abstract}



\section{INTRODUCTION}
\label{sec:intro}  

The nature of the dark energy that drives the accelerated expansion of the
Universe is one of the greatest mysteries in modern science. Of the
observational techniques that probe dark energy\cite{2013PhR...530...87W}, one
of the most promising is measuring the baryon acoustic oscillation (BAO) scale
using fluctuations in the large-scale distribution of neutral hydrogen.
\cite{2008PhRvL.100p1301L}

The BAO feature was imprinted in the matter correlation function at a scale of
approximately 150 comoving Mpc when baryons decoupled from radiation. By
measuring the BAO standard ruler in the large-scale structure across redshift,
the expansion history of the universe is measured.  In particular the period
from redshift  1--2 has the most power to distinguish between dark energy
models.  The first clear detection of BAO came from analyzing the Sloan Digital
Sky Survey (SDSS) luminous red galaxies at redshift
z$\sim$0.35\cite{2005ApJ...633..560E}.  The detection was verified in the two
degree field galaxy redshift survey at redshift
$\sim$0.2.\cite{2005MNRAS.362..505C}  It has more recently been measured at
three redshifts of $\sim$0.44, $\sim$0.6, and $\sim$0.7 in the WiggleZ Dark
Energy Survey\cite{2011MNRAS.418.1707B} and Baryon Oscillation Spectroscopic
Survey (BOSS) \cite{2014MNRAS.441...24A} at $z \sim 0.57$.  BAO were further
detected in the Ly-$\alpha$ forest at redshift $\sim$2.4 with
BOSS.\cite{2013JCAP...04..026S, 2013A&A...552A..96B}

The 21\,cm neutral hydrogen emission is an accurate tracer of matter on cosmological
scales. \cite{Chang:2010ds, 2013MNRAS.tmpL.125S, 2013ApJ...763L..20M} 
The isolation of the 21\,cm line eliminates the
spectral confusion handicapping optical surveys operating at redshifts of 1 to 3.  Since
the BAO angular scale is on the order of a degree on the sky, the detection of
individual galaxies as in an optical BAO survey is not necessary. Instead, a
21\,cm intensity map measures the aggregate 21\,cm emission at larger scales.  These features make neutral hydrogen an
excellent tracer of large scale structure and BAO at the redshifts where dark
energy becomes dominant.  

The properties of the instrument are set to measure the BAO scale.  The BAO first peak 
corresponds to an angular size of 1.35$^\circ$ at $z=2.5$ and 3$^\circ$ at $z=0.8$.  Nyquist sampling the BAO feature in a 21\,cm map
requires baseline lengths of 15\,m and 63\,m at the corresponding redshifts. Along
the line of sight, at redshift 0.8 (800\,MHz) the BAO scale corresponds to a
correlation at 20\,MHz separation, and at redshift 2.5 (400\,MHz) it corresponds to
a 12\,MHz separation correlation.

In order to measure the BAO in neutral hydrogen, we are building a close-packed
interferometer radio telescope,  the Canadian Hydrogen Intensity Mapping
Experiment (CHIME) Pathfinder. It is a smaller scale version of the full CHIME
instrument and is being built to inform the full CHIME design in addition to making
competitive measurements of the BAO scale.  The pathfinder is comprised of two
cylinders 20\,m wide by 37\,m long which are set 2\,m apart.  The cylinders are
oriented along the north-south direction and fixed.  The cylindrical
nature of the structure gives the instrument a $\sim$100$^\circ$ instantaneous field
of view in the north-south direction, allowing a full survey of half the
sky each sidereal day as the earth rotates. The close-packed configuration gives the instrument a high intensity mapping speed.

A summary of the Pathfinder specifications is in Table~\ref{tab:factsheet}. All 
components have been designed with an eye toward scaling the instrument by a 
factor of ten.  The
CHIME Pathfinder has 128 dual-polarization feeds and processes 400\,MHz of
bandwidth.  Scaling to the 1280 feeds required for the
CHIME telescope is made affordable today by advances in analog and digital
processing.  We have designed low-cost low-noise amplifiers based on
commercial components that achieve a
noise figure of 35\,K at room temperature.  This drastically simplifies the cylinder feed line
receiver design and number of components.  For the channelizer we use 
field-programmable gate array (FPGA) processors which offer many digital signal
processing blocks as well as many high-speed serial links, greatly
simplifying the design of the digital network required to re-arrange and
redistribute all the data for correlation. Our use of graphics processing units
(GPU) for the correlation allows for a rapid development cycle and unique additional processing.

\begin{table}[tb]
    \begin{center}
        \begin{tabular}{l|cc}
        \hline
        Frequency Range & 400\,MHz -- 800\,MHz \\
        \hline
        Redshift Range & 2.5 -- 0.8 \\
        \hline
        Beam Size &  1$^\circ$ -- 0.5$^\circ$    \\
        \hline
        Frequency Resolution & 390\,kHz, 1024 bins \\
        \hline
        E-W FoV & 2.5$^\circ$ -- 1.3$^\circ$ \\
        \hline
        N-S FoV & $\sim$100$^\circ$ \\
        \hline
        Single Source Observing Time per Day & Equator: 10\,min -- 5\,min \\
            & 45$^\circ$: 14\,min -- 7\,min\\
            & NCP: 24\,hr\\
        \hline
        Structure & 2 cylinders -- 37\,m x 20\,m \\
        \hline
        Number of Beams & 128 dual polarization \\
        \hline
        Receiver Noise Temperature & 50\,K \\
        \hline
        \end{tabular}
    \end{center}
    \caption{Pathfinder design parameters}
    \label{tab:factsheet}
\end{table}

The layout of the paper is as follows.  Section \ref{sec:science} describes
the science goals the CHIME Pathfinder hopes to attain.  Section
\ref{sec:telescope} describes the mechanical structure of the telescope,
Section \ref{sec:analog} describes the details of the feed and analog
electronics and Section \ref{sec:corr} describes the digital backend of the CHIME
Pathfinder.

\section{Science Goals}
\label{sec:science}

The CHIME Pathfinder will measure the large-scale structure of the Universe from
redshift 0.8--2.5 in 21\,cm emission.  Though the matter power spectrum contains
a wealth of useful information, we are primarily interested in measuring the
BAO, relics of primordial sound waves that produce enhanced correlation at a
characteristic scale. The physical observables corresponding to the BAO scale in
the radial (redshift) and transverse (angular) directions are $\Delta
z=H(z)r_s/c$ and $\Delta\theta=r_s/D_M(z)$, where $H(z)$ is the Hubble
parameter, $r_s$ is the sound horizon at baryon-photon decoupling, and $D_M(z)$
is the transverse comoving distance. The sound horizon is well-constrained by
Cosmic Microwave Background temperature anisotropy measurements
\cite{2013ApJS..208...19H, 2013arXiv1303.5076P}, allowing measurement of the BAO
scale as a function of redshift to directly probe the distance-redshift relation
and expansion history of the Universe.

A major challenge faced by hydrogen intensity mapping experiments is the
presence of bright astrophysical radio emission at the same frequencies as the
cosmological 21\,cm emission we are searching for. These foregrounds are dominated by
synchrotron emission from both our own galaxy and high redshift radio galaxies.
They have brightness temperatures up to 700\,K, many orders of
magnitude brighter than the 0.1\,mK signal we are looking for.

\begin{figure}[tb]
    \begin{center}
        \includegraphics[width=0.9\textwidth]{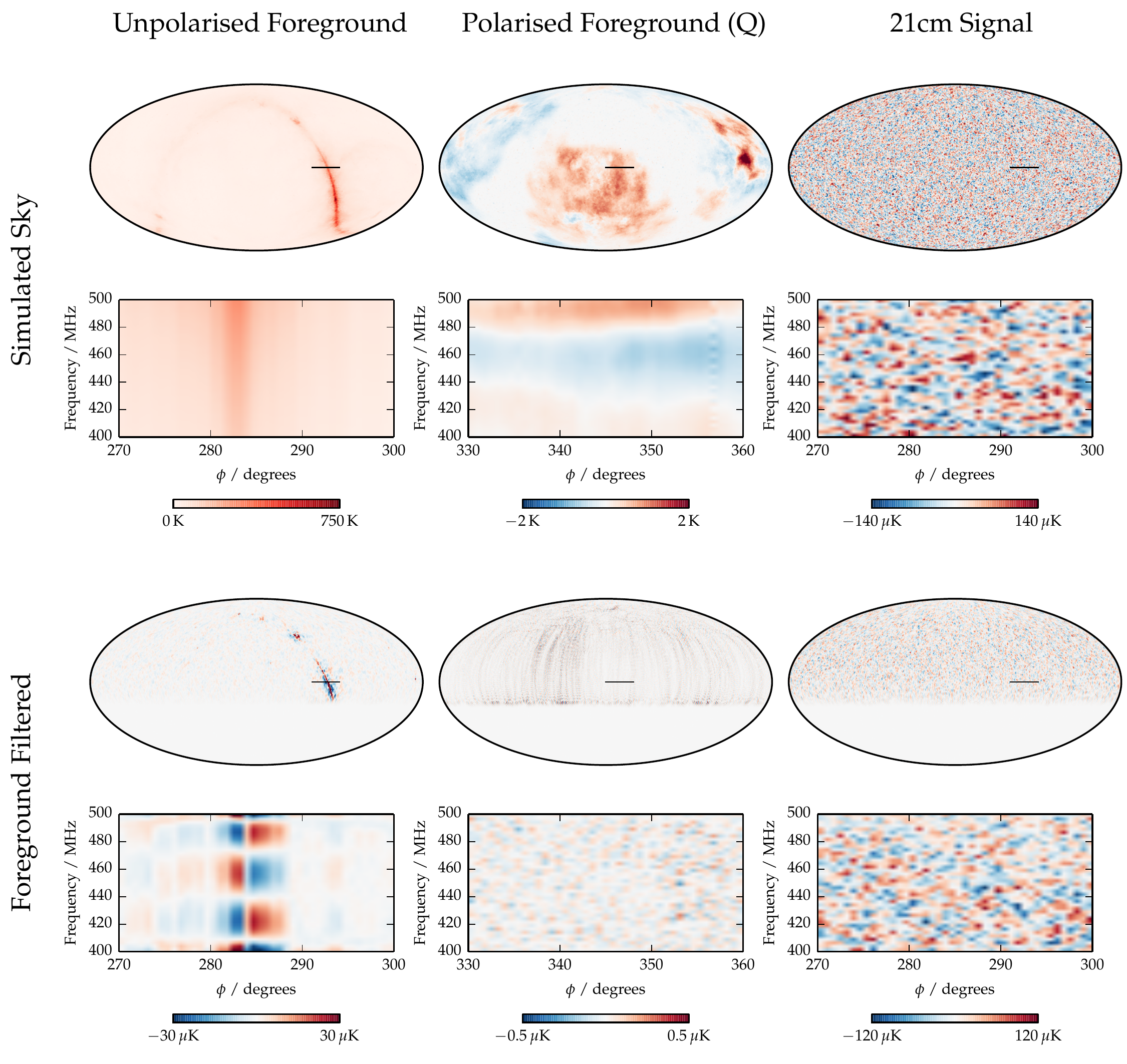}
    \end{center}
    \caption{This plot illustrates the process of foreground removal on 
    simulations of the radio sky. The top row of plots shows sky maps of the 
    individual components: unpolarized foregrounds, polarized foregrounds 
    (showing Stokes Q only), and the 21\,cm signal. On the bottom row we show the
    maps we would make after foreground cleaning visibilities from the CHIME 
    Pathfinder.  In each, the upper panel shows a frequency slice at 400\,MHz, 
    and the lower panel a slice through the galactic plane from
    400 to 500\,MHz.  Both the polarized and unpolarized foregrounds become
    substantially suppressed, whereas the 21\,cm signal is largely unaffected. 
    This leaves a clear correspondence between the original signal simulation 
    and the foreground subtracted signal, while leaving the foreground 
    residuals over 10 times smaller in amplitude than the 
    signal. Figure from Shaw et. al. 2014.\cite{Shaw:2014vy}}
    \label{fig:fgremoval}
\end{figure}

Fortunately the synchrotron emission is spectrally very smooth, and can be
separated on this basis from the large-scale structure 21\,cm signal. Though this is
superficially straightforward, instrumental realities such as 
frequency-dependent beams and polarization leakage present added challenges, which can be
surmounted with precise calibration \cite{newburgh2014} and new analysis
techniques \cite{Shaw:2014vy}. In Figure~\ref{fig:fgremoval} we demonstrate the
analysis of simulated time stream data, showing how we can effectively remove
foregrounds using a Karhunen-Lo\`eve transform to separate the signals based on
their statistics.

The primary quantity we are interested in measuring is the spatial power
spectrum of the large-scale structure, which contains most of the useful
cosmological information. Unfortunately the removal of foreground contamination
inevitably reduces our sensitivity to the signal we are interested in. Using the
\textit{m}-mode formalism, the two-dimensional matter power spectrum can be reconstructed
in the presence of foregrounds\cite{2014ApJ...781...57S, Shaw:2014vy}. As expected
we lose sensitivity to the largest scale fluctuations in the line-of-sight
direction, but still retain significant sensitivity to the peaks of the BAO.

\begin{figure}[tb]
    \begin{center}
        \includegraphics[width=0.6\textwidth]{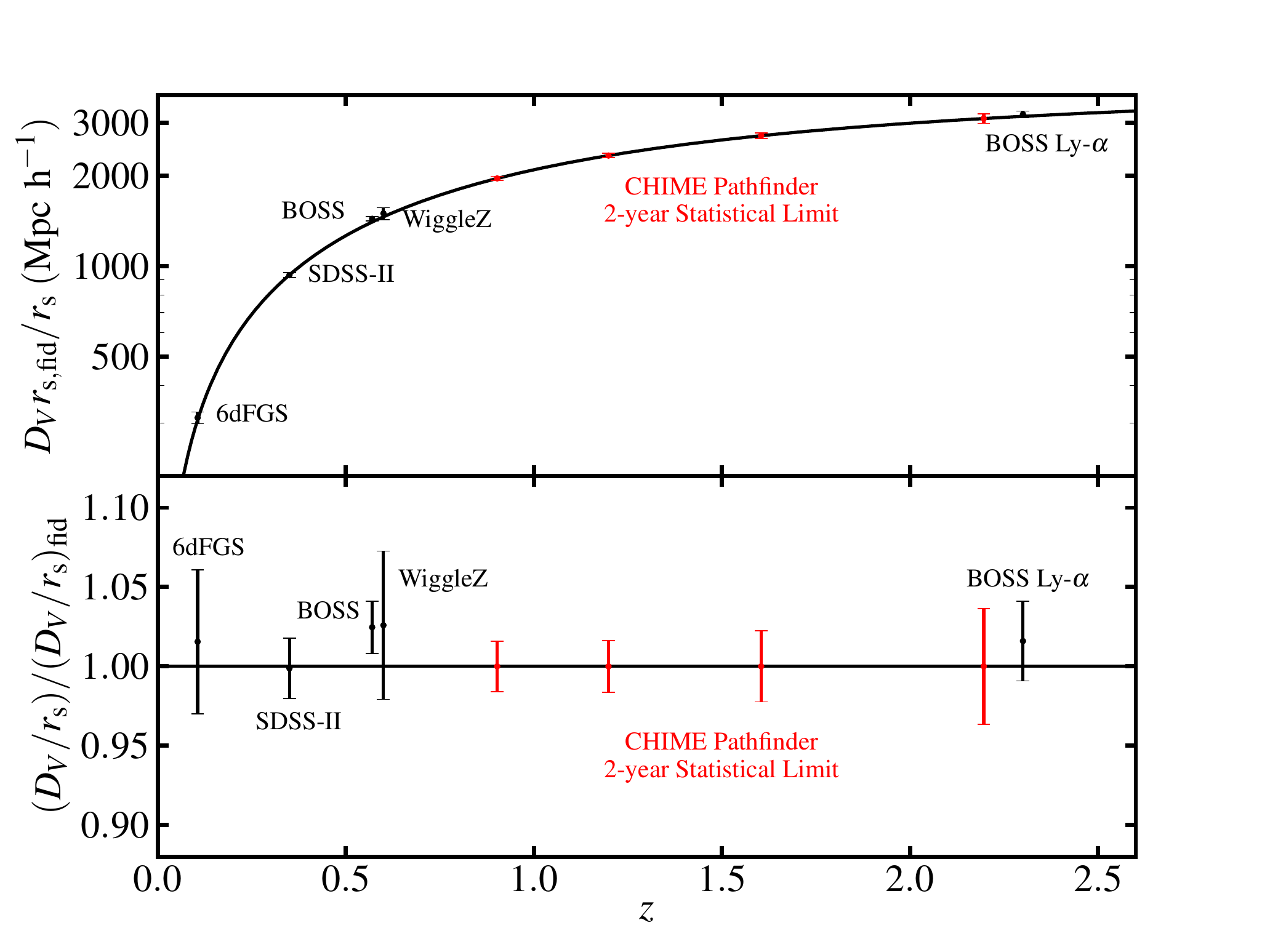}
    \end{center}
    \caption{Forecast CHIME Pathfinder constraints on the expansion history parameterized using the ratio of $D_V$ (see text) to the sound horizon $r_s$ as a function of redshift,
    shown relative to a fiducial $\Lambda$CDM cosmology with $h=0.7$, 
    $\Omega_\Lambda = 0.7$, $\Omega_m = 0.3$. Also plotted are constraints from 
    the 6dFGS\cite{2005MNRAS.362..505C}, SDSS \cite{2005ApJ...633..560E}, 
    BOSS \cite{2014MNRAS.441...24A}, WiggleZ \cite{2011MNRAS.418.1707B} and the 
    BOSS Ly-$\alpha$ results \cite{2013A&A...552A..96B}.  For the BOSS 
    Ly-$\alpha$ results the error bars are just a direct fractional error 
    translation from the $\alpha_{iso}$ constraints for illustrative purposes,
    and should not be interpreted as their constraint on $D_V$.    
    For the CHIME Pathfinder, the forecast error bars were calculated using the methods
    from Shaw et. al. 2014\cite{Shaw:2014vy} and represent the statistical limit with two years integration time.  }
    \label{fig:DA_H0}
\end{figure}

In Figure~\ref{fig:DA_H0} we illustrate the forecast constraints
that the CHIME Pathfinder could place on the expansion history though measuring
the BAO. The figure illustrates the forecast Pathfinder statistical limit of
$D_V(z)=\left[D_M(z)^2\frac{cz}{H(z)}\right]^{1/3}$ with two years observation time in comparison to current
measurements. 

\subsection{Ancillary Science}

The CHIME Pathfinder is an excellent platform to pursue ancillary science
goals due to its design and operating parameters. Some of those potential
goals are listed below:

Pulsars:  The CHIME Pathfinder's wide field of view lends itself to the
monitoring of pulsar timing.  The telescope is able to see every pulsar
in the northern hemisphere for at least $\sim$5 minutes each day.  Every day
hundreds of pulsars can be observed and their dispersion measure, timing
and scintillation can be monitored.  An additional backend is being built
which will receive full rate data on individual sky beams for measurement 
of pulse arrival times for a list of known pulsars.

Radio transients:  The survey nature of the CHIME Pathfinder also lends itself
to serendipitous detection of radio transients,  including fast radio bursts.
\cite{2007Sci...318..777L, 2013Sci...341...53T, 2014arXiv1404.2934S}
The CHIME Pathfinder has the sensitivity and field of view to discover many
undetected transient sources.  An exciting possible transient source is the radio afterglow from compact binary coalescence.\cite{Feng:2014uf}

Galactic polarization and Rotation Measure:  In order to make a precise intensity map of the sky, the
CHIME Pathfinder must be able to remove the foreground polarized emission from
the galaxy.  The CHIME Pathfinder will produce an accurate map of foreground
polarization, which can be used to further understand Galactic magnetic fields.

Cosmic rays:  The close-packed nature of the CHIME Pathfinder will allow for the
telescope to detect the radio emission from cosmic-ray air showers.  Information
about the air shower can be reconstructed from the arrival time, intensity, and
polarization of the air-shower pulse.

\section{Telescope}
\label{sec:telescope}
\begin{figure}[htbp]
\centering{
\subfloat[]{\includegraphics[height=1.6in]{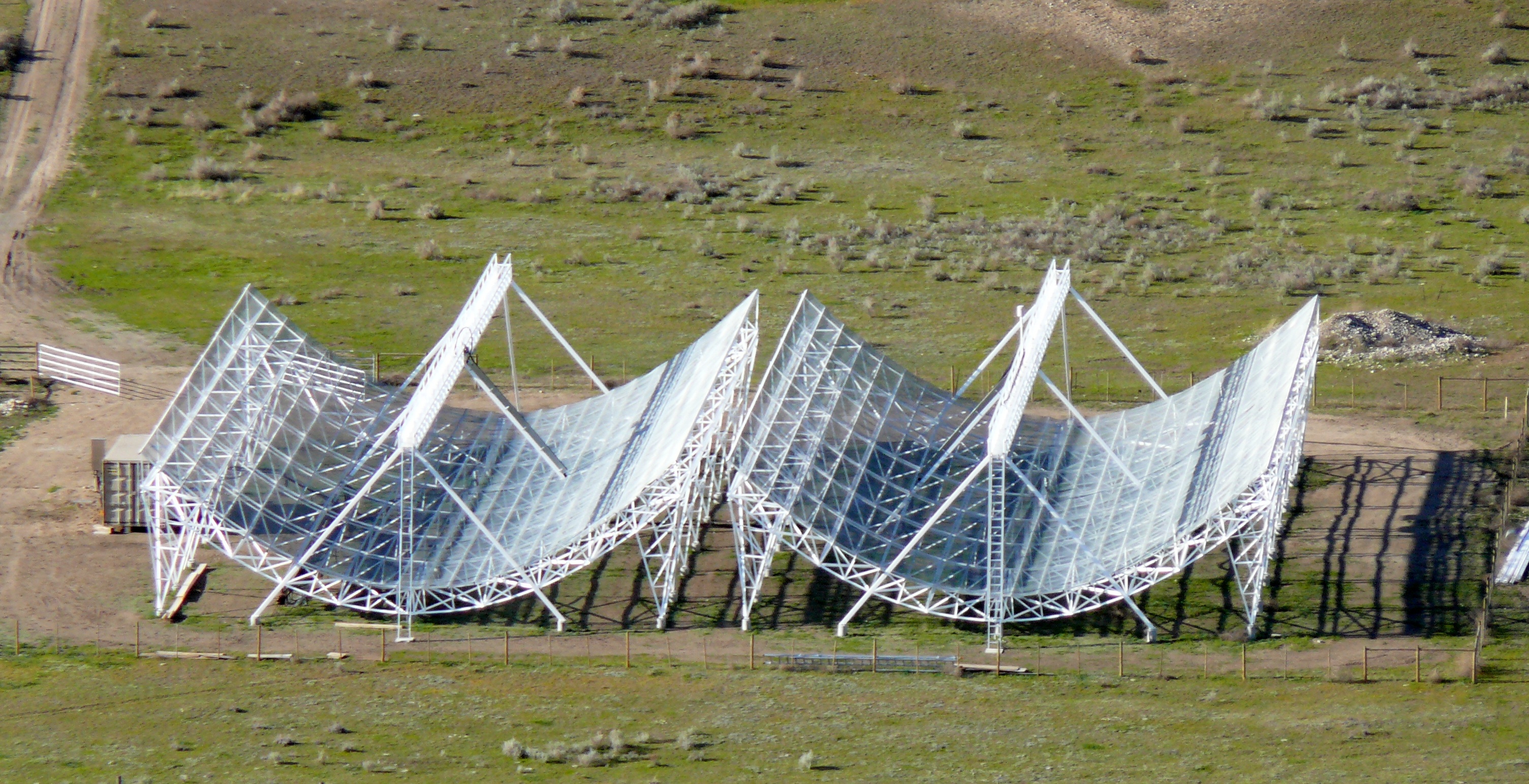}
\label{fig:cylImage}} 
\subfloat[]{\includegraphics[height=1.8in]{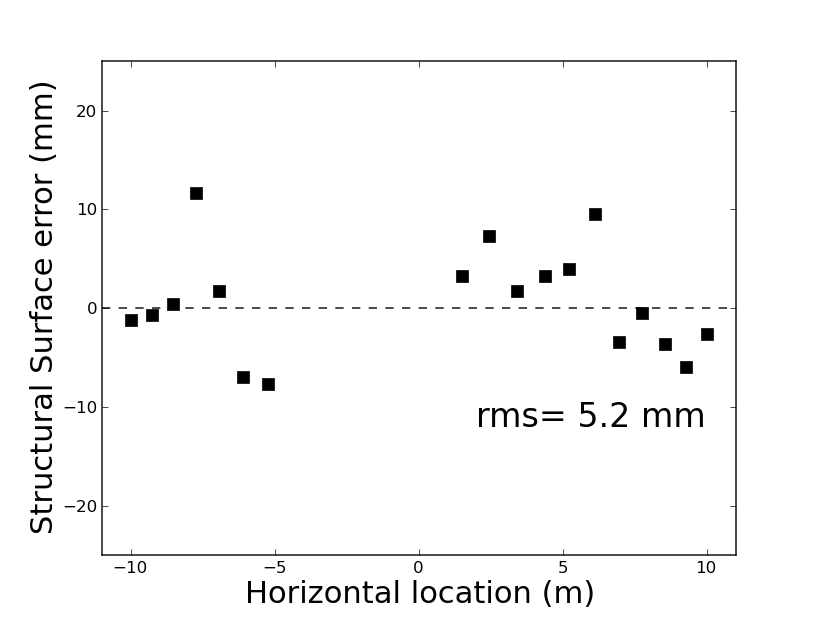}
\label{fig:cylAccuracy}} \\
\subfloat[]{\includegraphics[height=1.35in]{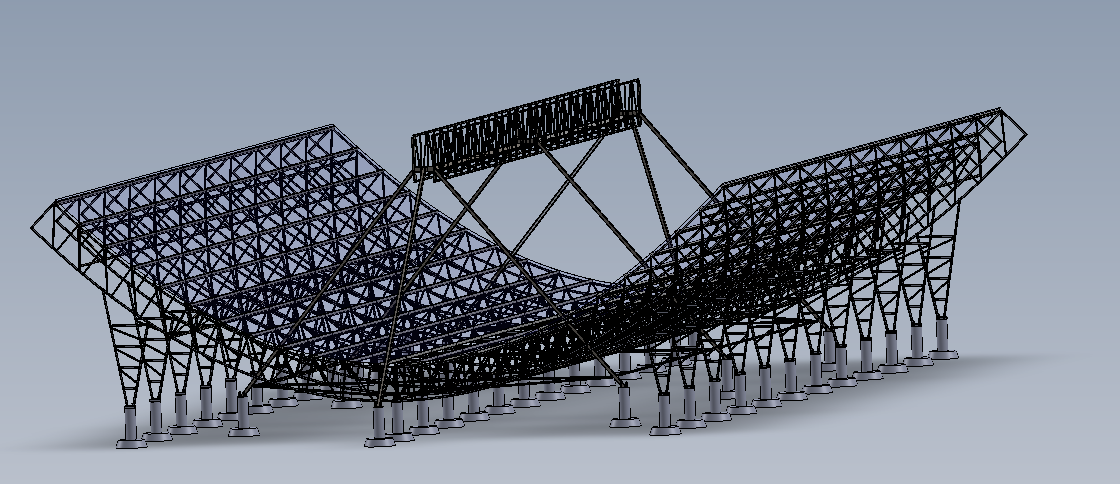}
\label{fig:cylStructure}} 
\subfloat[]{\includegraphics[height=1.8in]{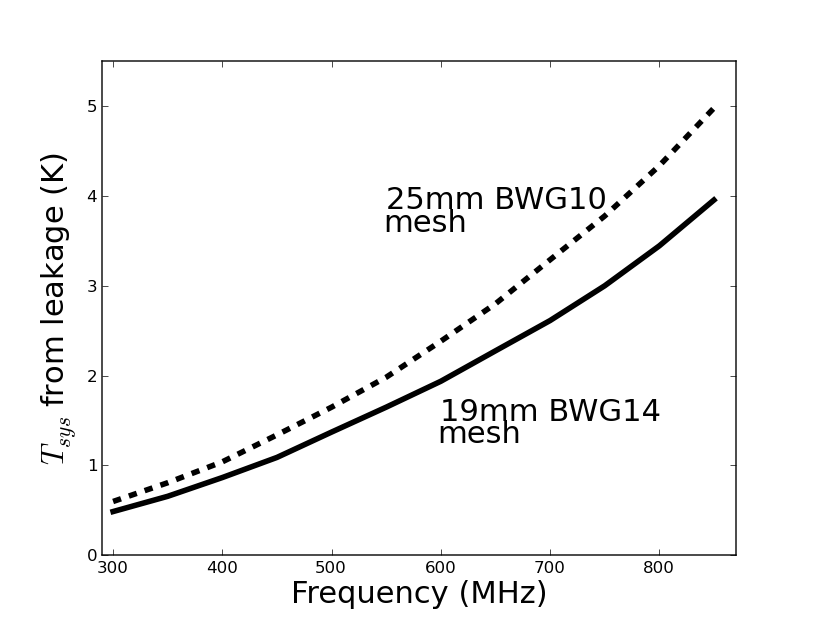}
\label{fig:cylMesh}} 
}
\caption{(a)  Photograph of the CHIME Pathfinder at the Dominion Radio 
Astrophysical Observatory (DRAO), April
2014. The digital correlator is housed in an RFI enclosure built into
the sea container visible at the left hand (western) edge of the
western cylinder.  Signals from the feeds are brought down the
middle diagonal support legs and across the the correlator under the
reflectors.  Removable ladders visible at the south end of each focal line allow
access. 
(b) Steel support structure normal surface error of the
western cylinder obtained from photogrammetry is plotted in
millimeters.  The quantity plotted is deviation of the supports from the design shape,
not from a best fit parabola.  The mesh has been measured to have a
shape deviation of 14\,mm rms with respect to the steel
support structures for the surface mesh. 
(c)  A CAD model of the telescope 
support structure.  The cement
base of each support sits 70cm below grade to avoid possible
frost heave. Each truss is supported on three legs.∫ Each focal line
sits on seven legs, with the northern leg only serving to lock out a
north-south parallelogram motion of the six other legs.  The support
legs are 10\,cm x 15\,cm hollow steel structures and pass above the
instrument ground plane, allowing the one-dimensional feed array to
run uninterrupted along the focal line. 
(d) Frequency-dependent noise leakage of 
an assumed 300\,K load
at normal incidence.  The calculations have been performed for mesh
with 19\,mm spacing made of BWG 14 wire and 25\,mm spacing, BWG 10 wire.
These are the heaviest wire gauges we could find in mesh of these
spacings.   The comparatively dry snow
falling at 590\,m elevation in the BC interior typically falls through
the 19\,mm mesh from which CHIME is built. }
\label{fig:cylinder}
\end{figure}

The CHIME telescope structure consists of two adjacent parabolic cylinders, with a ground plane, feeds, and low noise amplifiers held
along the focal line of each cylinder beneath a walkway, see Figure
\ref{fig:cylImage}.  The reflective surface is galvanized steel mesh
bolted to the underlying structure.  The cylinder axes run north-south.

Each cylinder is 20\,m across and 37\,m long, see Figure~\ref{fig:cylStructure}.
The focal length of the parabola is 5\,m  and the telescopes have a focal ratio
of f/0.25. The telescope was designed to rely on standard steel construction
practices as much as possible.  On-site welding is held to an absolute minimum
to avoid RF disturbance of the host observatory.  The structure consists of
parabolic steel trusses set every 5\,m north-south supported by cement feet which
bear below the frost line.  Purlins spaced 1\,m apart run parallel to the
cylinder axis, bolted to each truss.  The structure is not very different from
that of any warehouse roof, except that the east-west cross section is
parabolic, the roof itself is instead mesh, and there is no floor.  The
structure is designed to have a peak deformation of less than 2\,cm under a wind
and snow load of 1\,kN/m$^2$.

The mesh properties have been chosen as a compromise between reflectivity to radio wave and the ability to shed snow.   CHIME has no moving parts and so, unlike
a conventional radio telescope, it can not be tipped over to shed accumulated
snowfall. A course mesh allows snow to fall through, which avoids loss of
observation time due to perturbed reflectivity.   However, a course mesh also
transmits more thermal radiation from the ground, raising the system
temperature.  The noise level from leakage of 300 K radiation through
mesh is plotted in Figure~\ref{fig:cylMesh} across the CHIME band for two
different mesh options. The CHIME Pathfinder uses 19 mm (3/4 in) spacing 14
Birmingham Wire Gauge (BWG) mesh, which is compared to a 25 mm (1 in) 10 BWG
mesh. These are the heaviest wire sizes available at each spacing and thus the
most reflective mesh at each size. The noise contribution of the 19 mm mesh is
roughly 1 K lower across the upper half of the CHIME passband, which corresponds
to a sensitivity difference per year equivalent to 15 days of extra
observations.  Over the 2013-14 winter snow was on the CHIME mesh only
immediately after the largest snow falls, for a total of four days. The mesh is
available in rolls of 2 m width. The 1m purlin spacing is chosen so that the
mesh sits on three purlins and takes the shape of the reflector. The mesh is
riveted every 75 cm along the purlin.

Photogrammetry has been used to measure the shape of the steel support
structure.  The results are shown in Figure~\ref{fig:cylAccuracy} as
deformations with respect to the nominal design shape.  The rms
deformation here is 5.2\,mm.  Deformation of the mesh with respect to the steel
structure has been measured by hand at hundreds of points across the structure.
The mesh surface error is 1.4\,cm rms and dominates the telescope
surface error budget.  This rms gives a cylinder efficiency of 99\%
and 97\% at 400\,MHz and 800\,MHz respectively. 

\section{Analog Chain}
\label{sec:analog}

\begin{figure}[htbp]
    \begin{center}
\includegraphics[width=\textwidth]{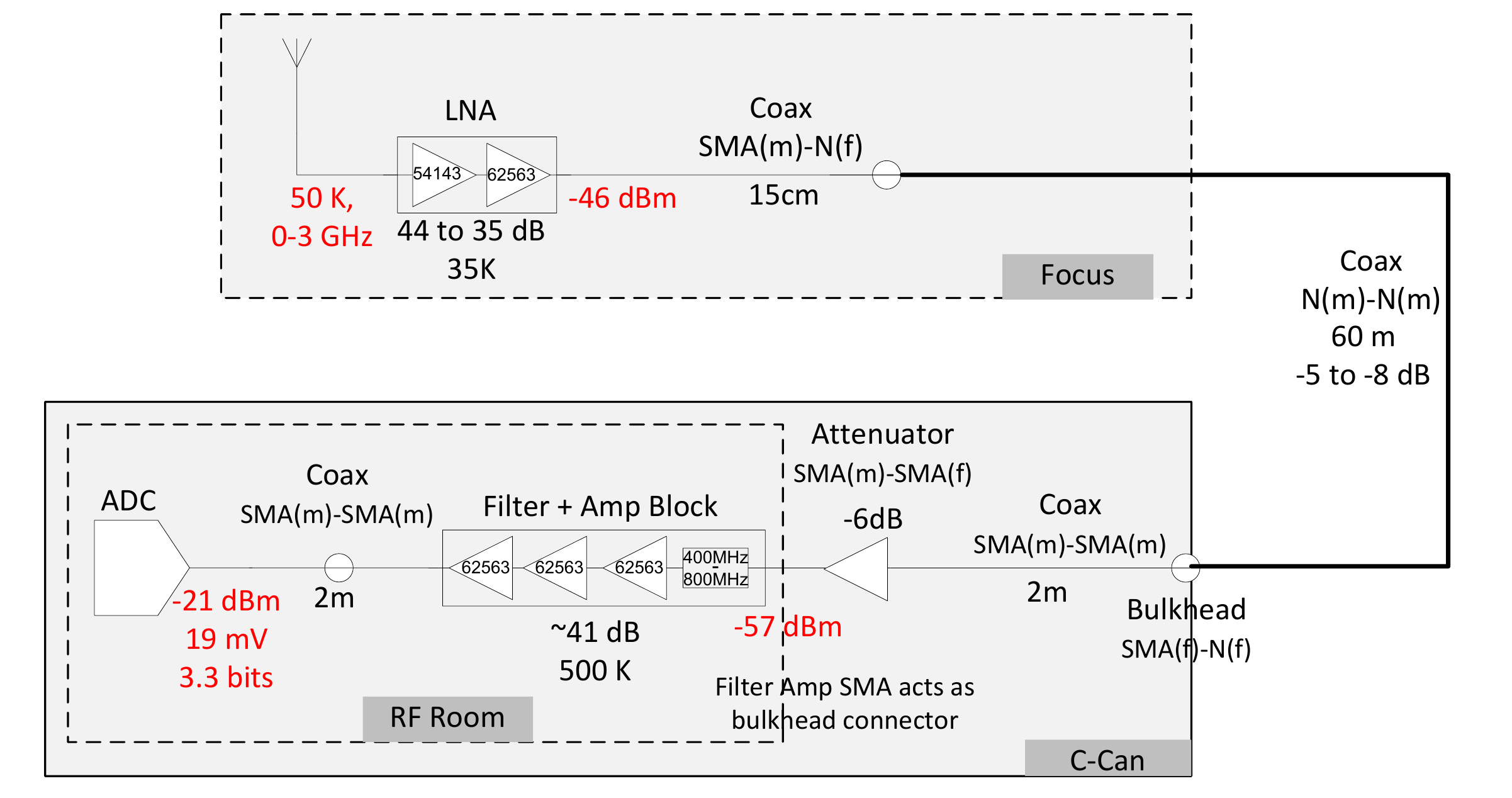}
    \end{center}

    \caption{The analog system signal chain overview.  The clover-leaf feed 
    receives the sky signal.  The received signal is then amplified by the 
    low-noise amplifier block by 35 to 44\,dB across the 400-800\,MHz band while 
    adding $\sim$35\,K noise.  The signal is then transmitted over 60\,m of LMR-400 
    coaxial cable to a central RF-shielded enclosure.  The signal is attenuated there by 
    approximately 6\,dB, customized to each amplifier chain.  The signal is then 
    bandpass filtered to 400-800\,MHz and further amplified by $\sim$41\,dB to 
    achieve an input to the ADC of $-$21\,dBm power.}
    
    \label{fig:analogChain}
    \vspace{0.2in}
\end{figure}

The analog system of the CHIME Pathfinder consists of the components shown in
Figure~\ref{fig:analogChain}.  The feeds and LNAs are located along the focal
line.  The signals are then sent over coaxial cables to a shielded RF enclosure.
The signals are then bandpass filtered and further amplified to achieve an input
power of -21\,dbm at the input of the ADCs.  The overall receiver temperature
design for CHIME is 50\,K.  This includes ground effects but does not contain
any contribution from the sky.  The analog components are described in more
detail in the sections below.

\subsection{Feed}

\begin{figure}[htbp]
\centering{
\subfloat[]{\includegraphics[height=1.9in]{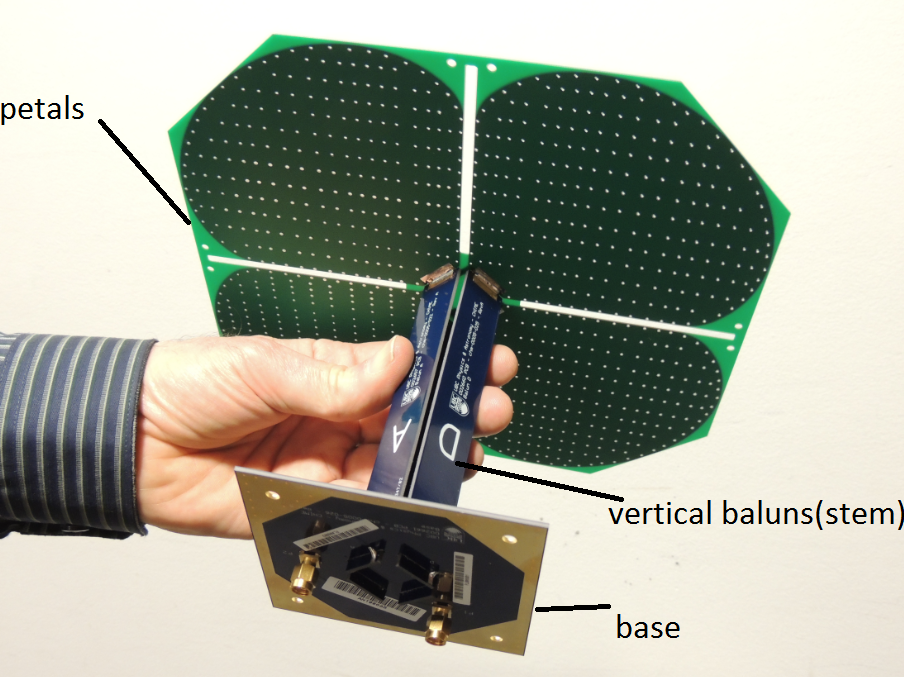}
\label{fig:feedgeoImg}}
\hspace{2mm}
\subfloat[]{\includegraphics[height=1.9in]{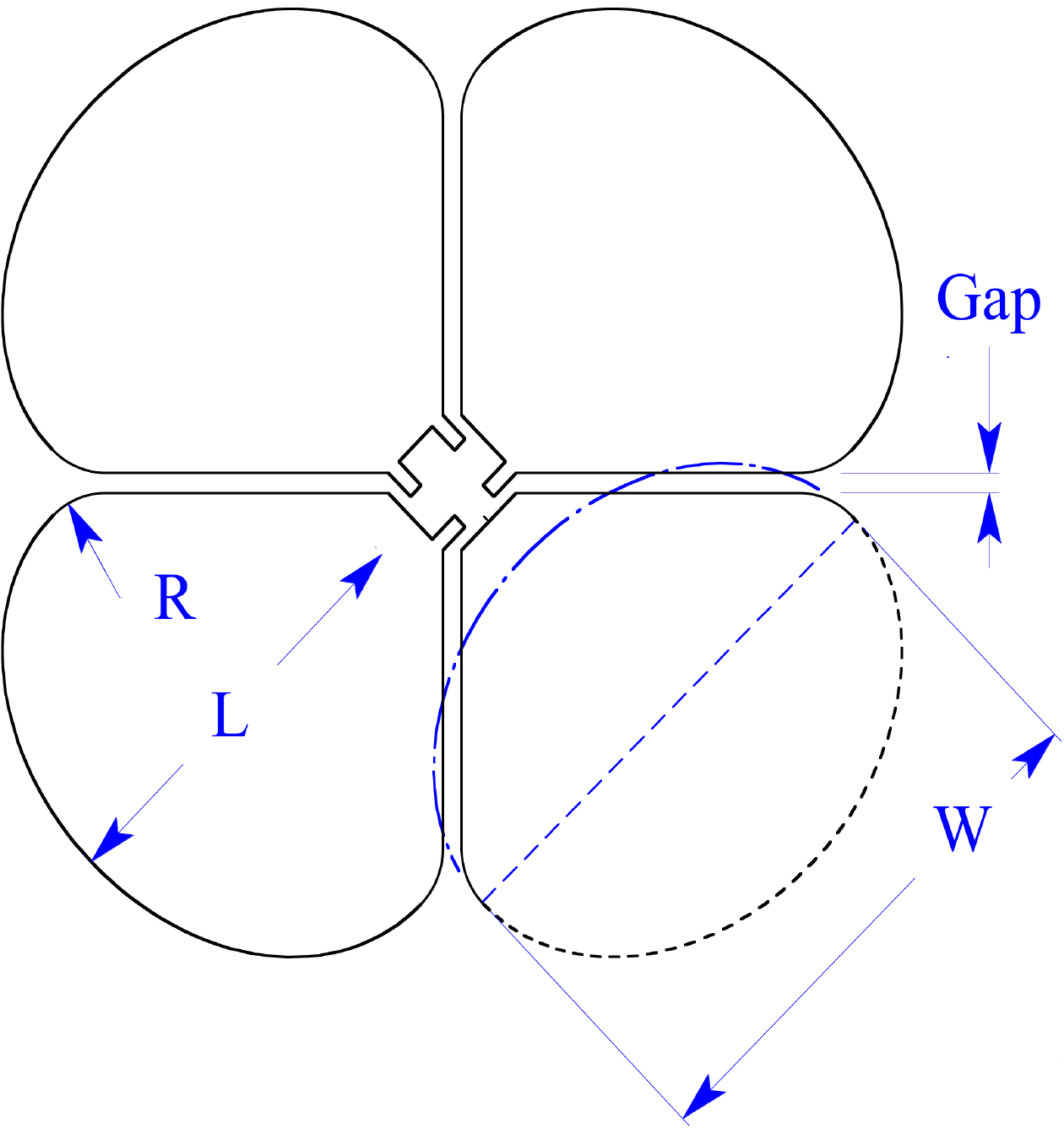}
\label{fig:feedgeoGeo}} 
\subfloat[]{\includegraphics[height=1.9in]{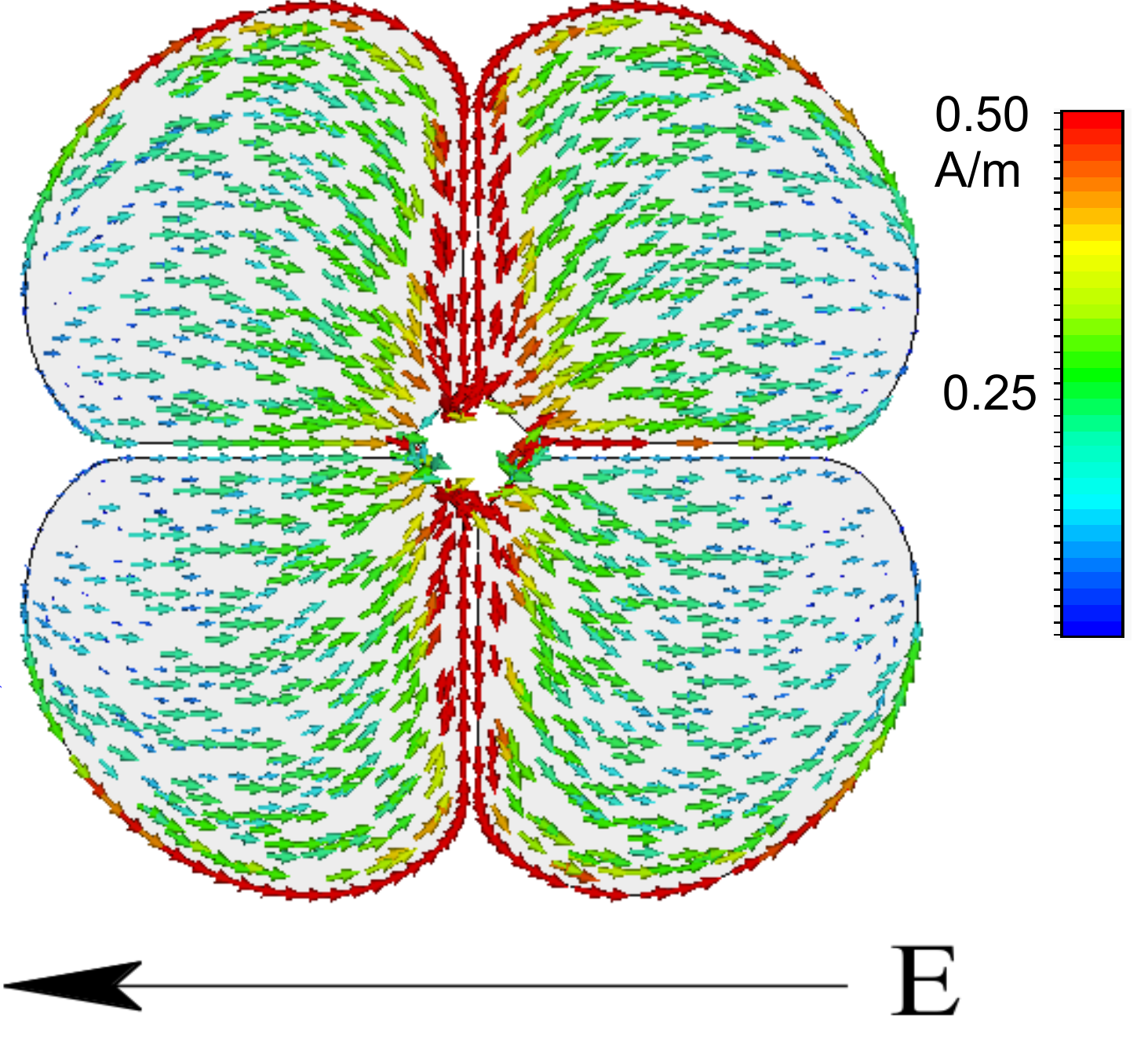}
\label{fig:feedgeoSim}} 
}
\caption{{(a) Photo of CHIME feed; (b) Geometry of petals with $W=138.5$\,mm,
$L=131.9$\,mm, $R=20$\,mm. This geometry is small enough that a feed element is
compatible with any azimuth orientation within the array.  (c)  The current pattern
from a CST simulation of the feed at 600\,MHz for the horizontal polarization
as indicated by the arrow labeled E.  }}
\label{feedgeo}
\end{figure}

The CHIME feed is a clover-leaf shaped compact dual-polarization feed. It
is a modification of the four-square feeds developed for the Molonglo
Observatory. \cite{Martin2008} The feed petals, balun stem, and support base are
all made from printed circuits boards (PCB), as shown in Figure~\ref{fig:feedgeoImg}.
As shown in Figure~\ref{fig:feedgeoGeo}, the petals have curved outer edges that
broaden the frequency response by reducing the number of individual resonant
dimensions. The curves are smooth and each petal is symmetric. The current
pattern from a CST studio\footnote{https://www.cst.com/} simulation is shown
in Figure~\ref{fig:feedgeoSim} for one linear polarization at 600\,MHz. The
currents near the gaps between petals run in opposing directions so they cancel,
and do not contribute to the radiation pattern. For this polarization, far-field
radiation arises from the coherent currents running along the curved outer edges
of the top and bottom pair of petals. \cite{deng2014}

Differential signals from pairs of adjacent
petals are combined through tuned baluns to form one single-ended output. Thus
each single polarization signal involves currents in all four petals.  Full baluns, from both polarizations, consist of four
identical microstrip transmission lines  along four vertical support
boards (the ``stem'') and a horizontal base board. Both of the single-ended outputs are
on the base board. Each transmission line is varied in several abrupt steps, and
the lengths and characteristic impedances of the transmission line segments are
carefully tuned to match impedance. Electrical losses in conventional circuit
board materials generate unacceptable loss for astronomical instrumentation, so
Teflon-based PCB is used for both the balun stem and support base.  Between
petals the PCB is completely removed to minimize loss along the slot transmission.

\begin{figure}[htbp]
\centering{
\subfloat[]{\includegraphics[width=.4\textwidth]{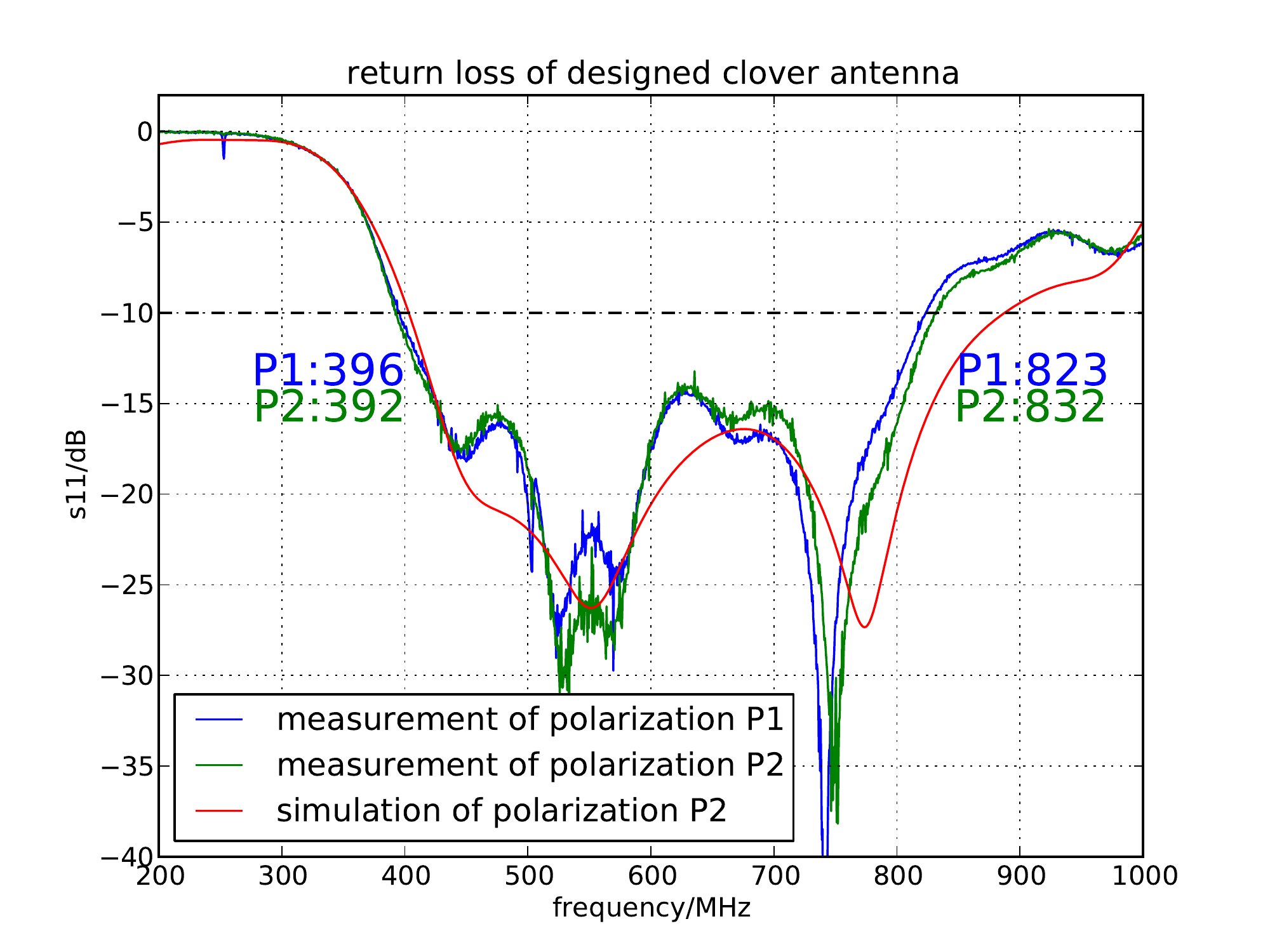}\label{feedresa}}
\subfloat[]{\includegraphics[width=.3\textwidth]{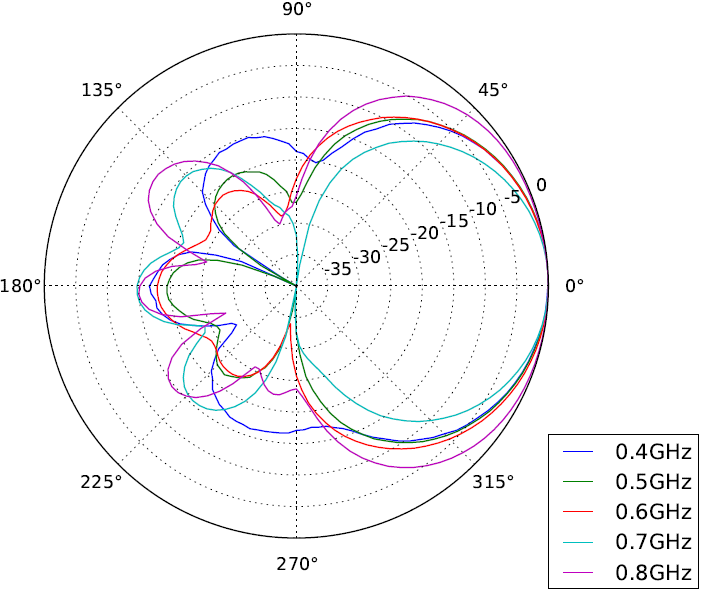}\label{feedresb}} 
\subfloat[]{\includegraphics[width=.3\textwidth]{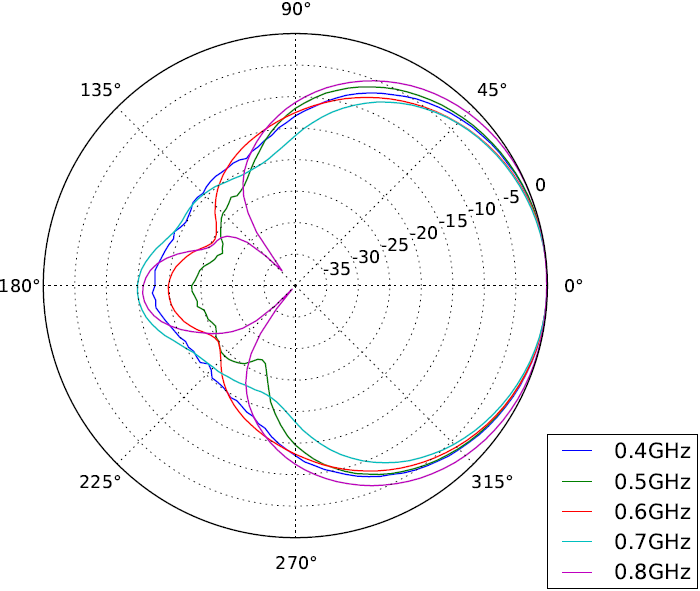}\label{feedresc}} 
}

\caption{(a) Measured return loss compared with simulation.  Note the 
similarity between two polarizations. (b) Measured E plane of polarization P2. 
(c) Measured H plane of polarization P2.
}

\end{figure}

The measured returned loss compared with simulation is shown in Figure~\ref{feedresa}, and the 
measured beam pattern is shown in Figures~\ref{feedresb}
and \ref{feedresc}. 

\subsection{LNA}

\begin{figure}[htb]
    \begin{center}
    \subfloat[]{\includegraphics[height=1.25in]{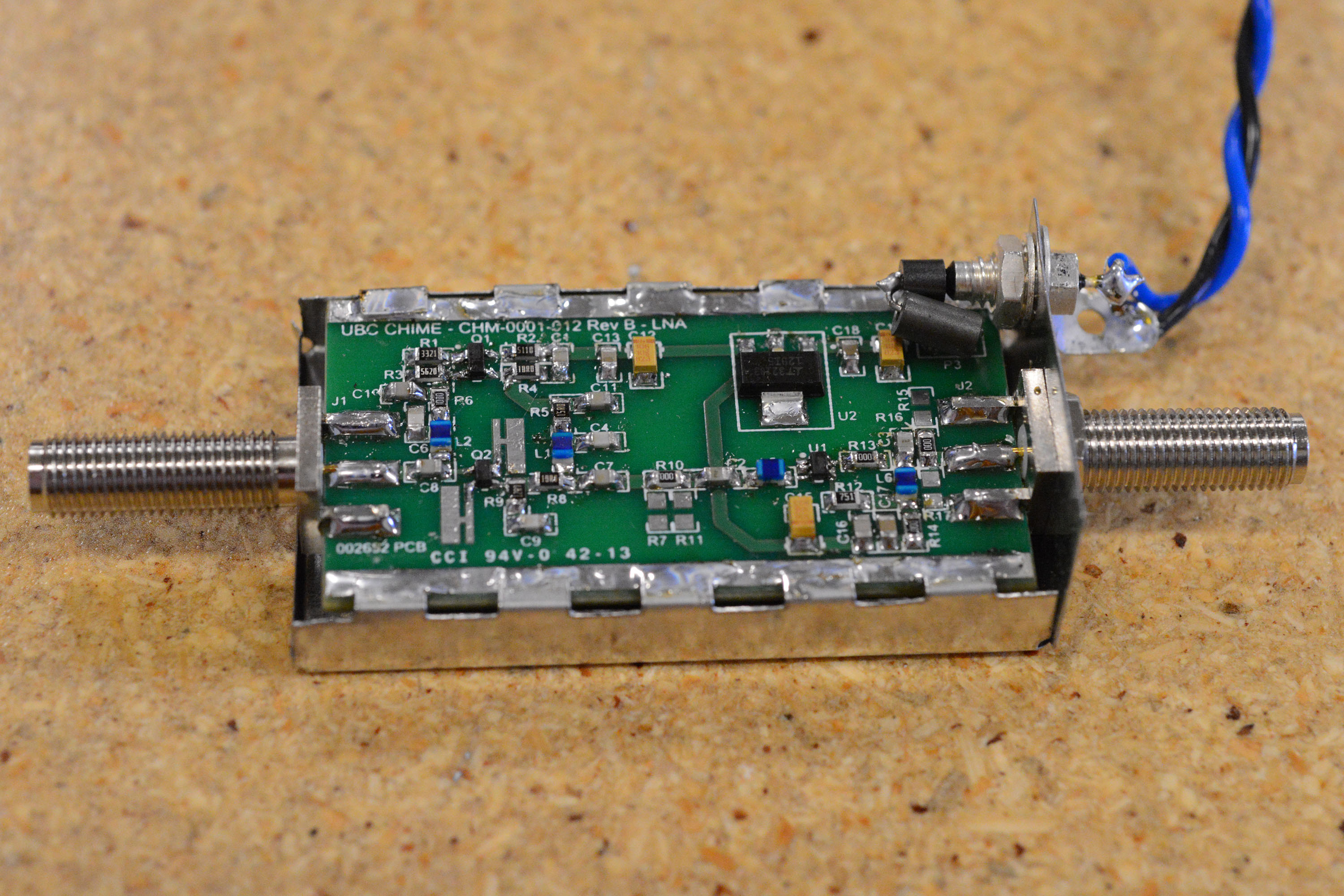}}
    \hspace{0.5in}
    \subfloat[]{
    \includegraphics[height=1.5in]{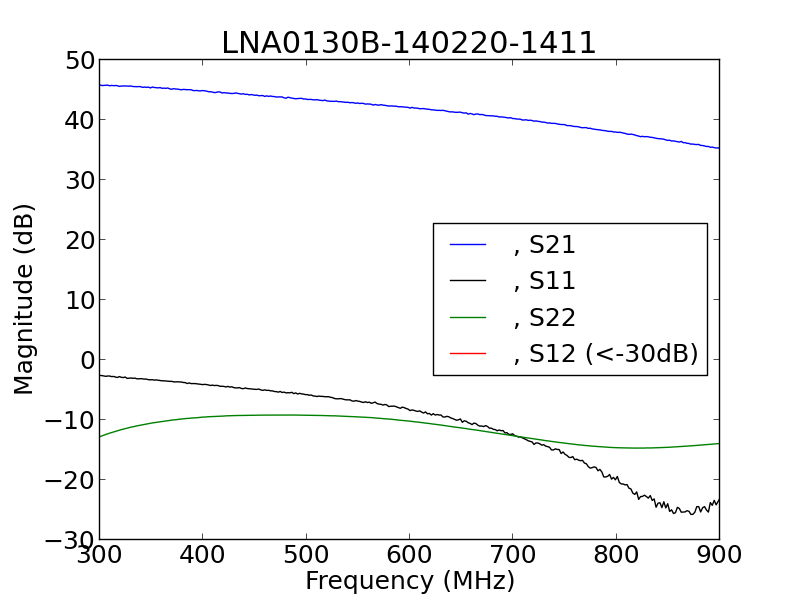}
    \label{fig:lnaSparam}
    } \\
    \subfloat[]{
    \includegraphics[height=2in]{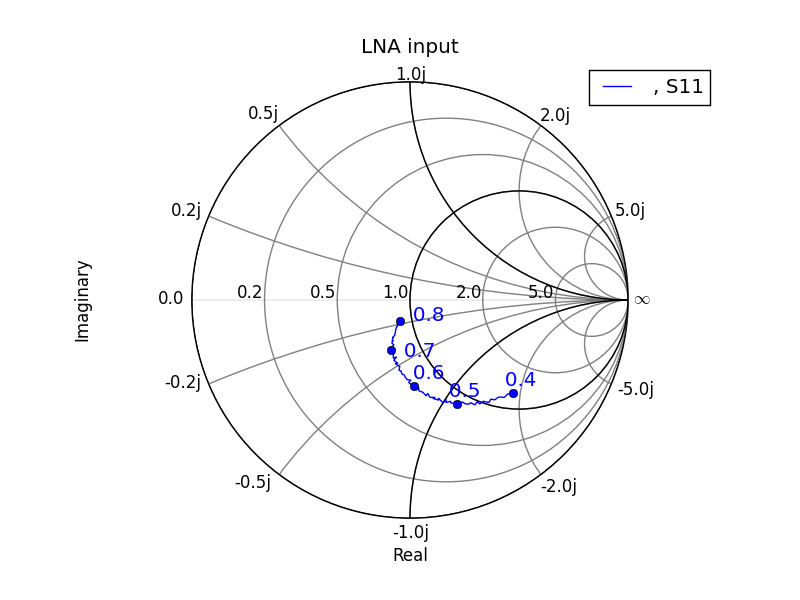}
    \label{fig:lnaSmith}
    }
    \subfloat[]{
    \includegraphics[height=2in]{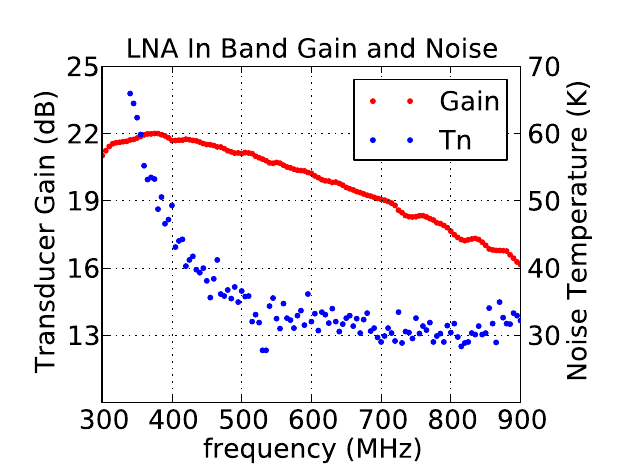}
    \label{fig:lnaNoise}
    }
    \end{center}
    \caption{(a) Image of the CHIME 
    LNA.  The circuit board is soldered to the case to reduce internal 
    resonances.  (b)  Measured S-parameters of the two-stage LNA. (c) Smith 
    Chart of the LNA input matching relative to 50 ohms.  
    (d) Measured gain and noise of the first stage of the LNA only. \vspace{0.1in}}
    \label{fig:lna}
\end{figure}

The low-noise amplifier (LNA) is located at the focus and is directly attached
to the feed.  The LNA has two gain stages.  The first is based on an Avago 54143
GaAs enhancement mode pseudomorphic high electron mobility transistor (E-pHEMT).
It is followed by an Avago MGA-62563 E-pHEMT radio frequency integrated circuit
(RFIC).  The achieved noise figure with this design is 35\,K across most of the
band, as shown in Figure~\ref{fig:lnaNoise}. 

The input matching, output matching and feedback of the amplifier were all
designed to achieve a low noise figure. The resulting gain and matching
S-parameters are shown in Figure~\ref{fig:lnaSparam}.  The input has a
relatively high reflection coefficient as the amplifier is noise matched instead
of impedance matched. This does have the effect of reflecting more than ten
percent of the incoming power back out of the feed. 

\subsection{Filter Amplifier}

\begin{figure}[htb]
    \begin{center}
    \subfloat[]{
    \begin{minipage}[c][1\width]{0.5\textwidth}
    \centering
    \includegraphics[width=0.9\textwidth]{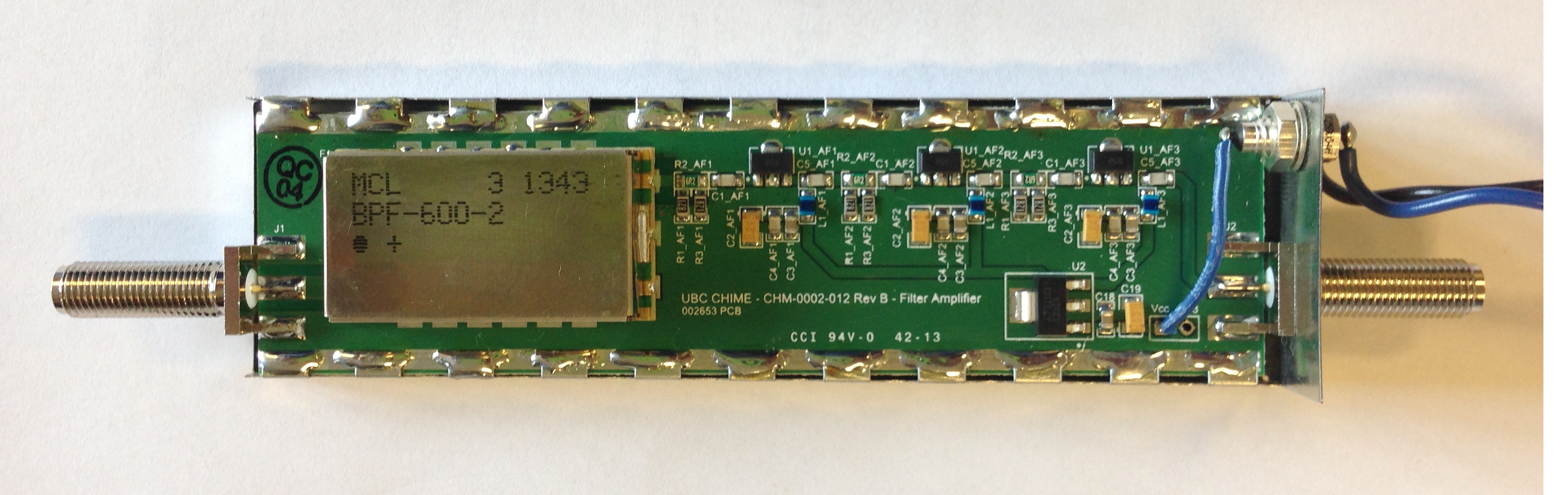} 
    \vspace{-1in}
    \end{minipage}}
    \subfloat[]{
    \begin{minipage}[c][1\width]{0.5\textwidth}
    \centering
    \includegraphics[width=0.9\textwidth]{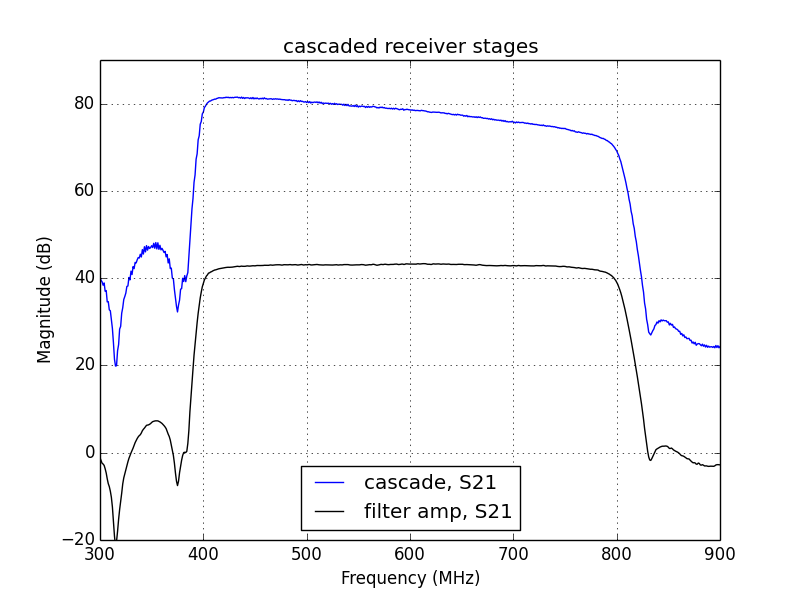} 
    \vspace{-1in}
    \end{minipage}}
    \end{center}
    \caption{ 
    (a) Image of the CHIME amplifier and band-defining filter. Input on the left. (b) Gain and passband of the filter-amplifier block labeled filter amp plotted along with the full analog chain labeled cascade. The passband of the filter amplifier block is designed to be very flat with frequency. The entire analog chain has a slope from low to high frequency primarily due to the LNA gain and analog cabling.  }
    \label{fig:fla}
\end{figure}

A 60\,m coaxial cable connects the ouput of each LNA to the sea container where
each signal is received by a filter amplifier. This block comprises a  
custom-made  Minicircuits band-pass 400-800\,MHz filter followed by 3 stages of  gain.
The S-parameters for this block are shown in Figure~\ref{fig:fla}.  The amplifier
has a flat passband with less than 3 dB variation.  It has greater than 20\,dB
rejection by 390\,MHz on the low side and 815\,MHz on the high side.  It is a highly linear device, with a measured output power compression point of 29\,dBm.  

\section{Digital Backend}
\label{sec:corr}

\begin{figure}[htb]
    \begin{center}
    \subfloat[]{
        \includegraphics[width=0.7\textwidth]{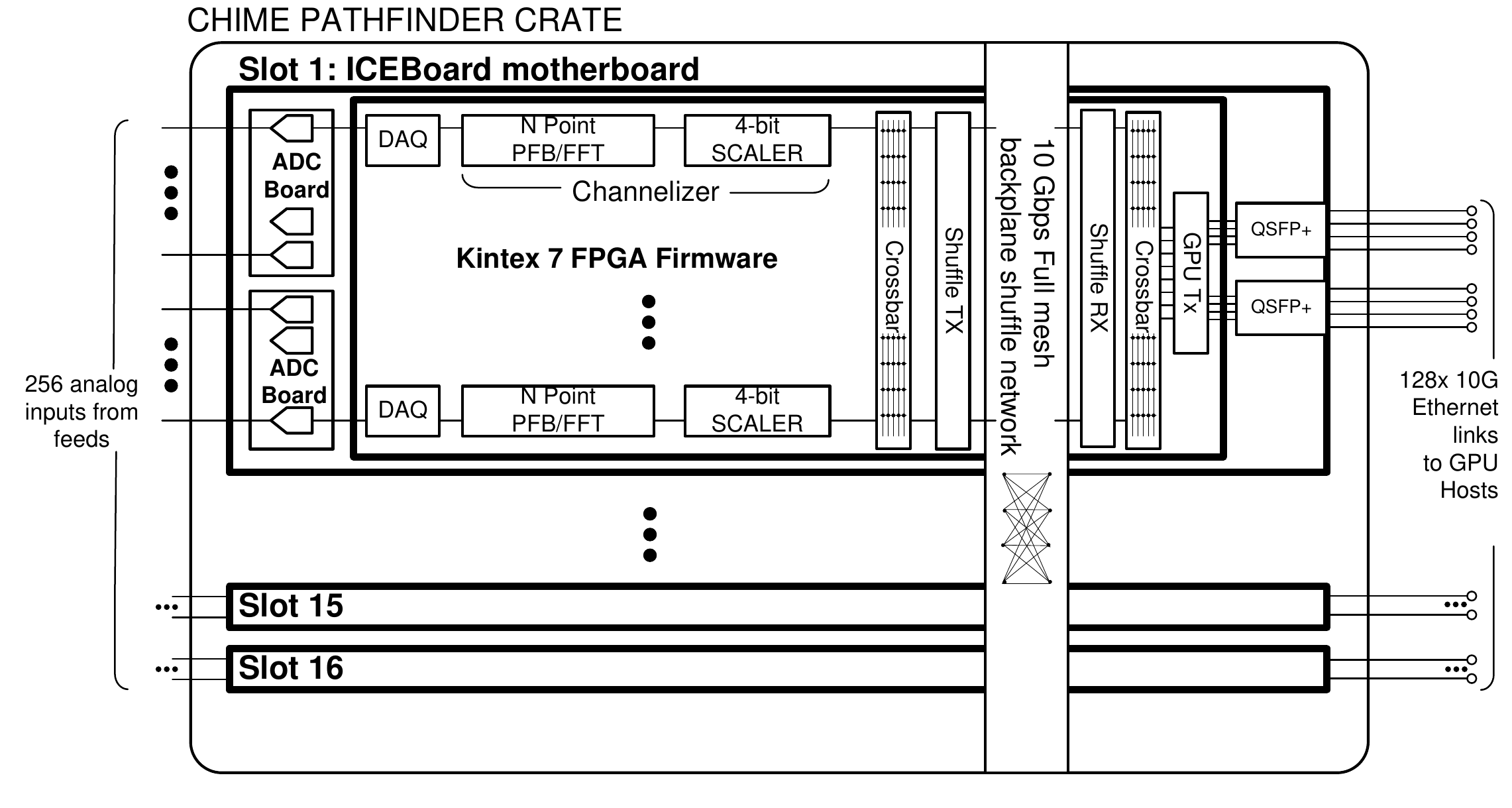} \label{fig:fpgaOverview}
        }\\
    \subfloat[]{
        \includegraphics[width=0.5\textwidth]{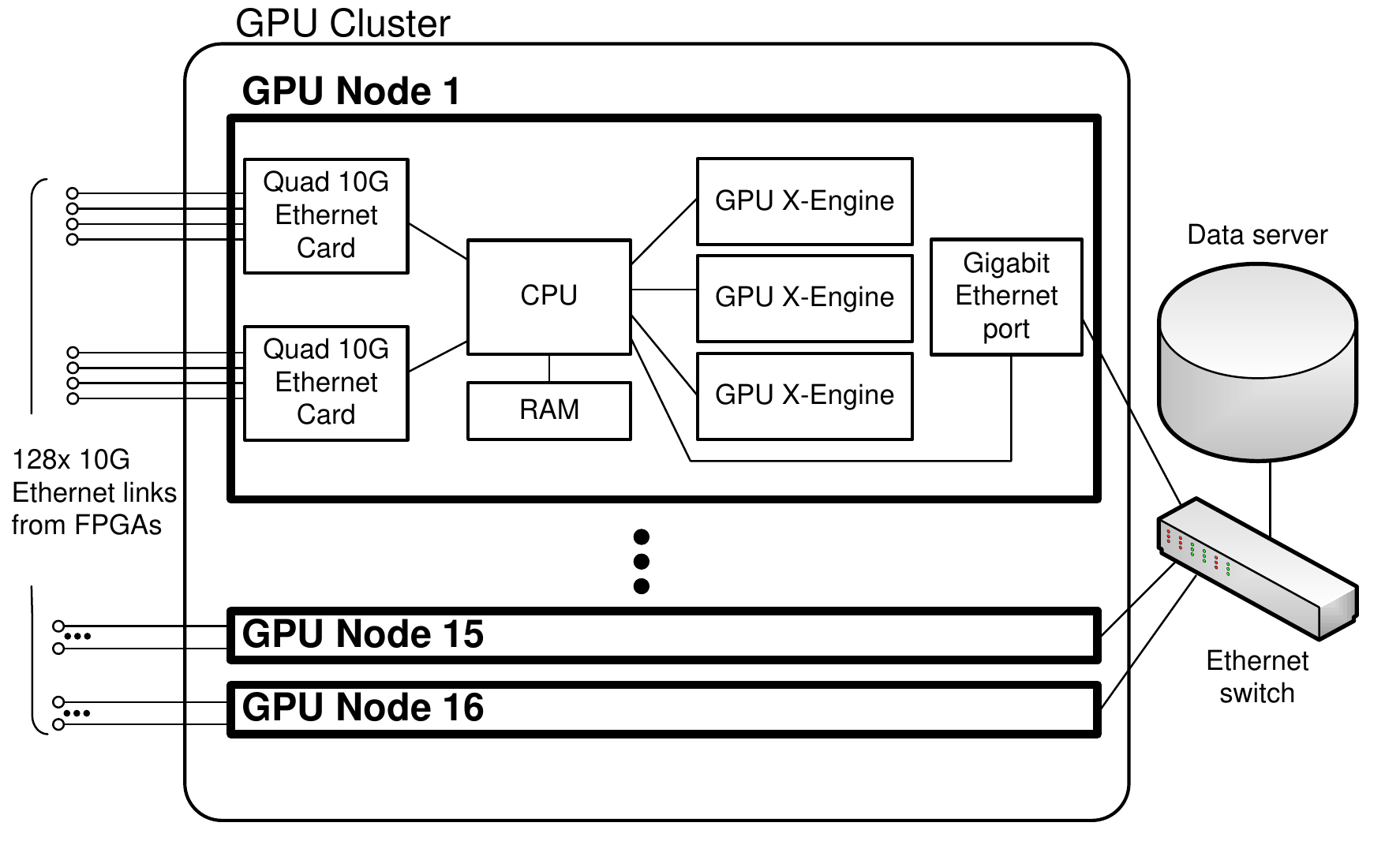}\label{fig:gpuBlock}
        }\\
    \end{center}
    \caption{Digital backend system overview.  a) Data is sampled by ADCs and is acquired and framed by the FPGA DAQ module, and is fed into the channelizer (F-engine) which performs a PFB/FFT and scales the data back into (4+4) bit values.  The data is rearranged in the
    FPGA and through a high-speed backplane network and sent over 10 gigabit
    Ethernet to a GPU farm to be correlated. b) Diagram of Pathfinder X-engine
    GPU system. Data is received by two quad 10 gigabit Ethernet cards and
    passed into system memory.  The data is then transferred in blocks to the
    GPU X-engine kernel for multiplication and accumulation.  Finally the
    correlated signal is offloaded to a data server for storage and further
    processing. }
    \label{fig:digOverview}
\end{figure}

The digital backend of the CHIME Pathfinder takes the overall structure of an FX correlator and is implemented into the main
components shown in Figure~\ref{fig:digOverview}:

\begin {itemize}      
    \item  The analog-to-digital converters that sample the sky signal are located on daughter cards
that attach to custom FPGA motherboards.
    \item The channelizer (F-engine) is implemented in each motherboard's FPGA to split
the 400\,MHz-wide analog signals into 1024 frequency bins 390\,kHz wide.
    \item The crossbar and shuffle modules re-organize the channelized data from all the motherboards
in a crate in order to concentrate the data for a subset of frequencies into a single FPGA. A 10 Gbps full-mesh network connecting every motherboard is implemented using a passive custom backplane driven by high speed serial transceivers on the FPGA.
    \item The offload link packetizes the shuffled data into 10 gigabit Ethernet packets and streams those packets to the GPU correlator host computers. 
    \item The correlator (X-engine) is implemented in a dedicated computing cluster, where each node receives the data with 10 gigabit Ethernet inputs, processes the packet header information and moves the data to system memory.
    \item  The GPUs are used to perform efficient, real-time full $N^2$ correlation and averaging of the data. 
    \item Commodity gigabit ethernet switches are used to collect the
      data onto a server which stores the integrated data. A server on
      this same network is used to configure and monitor the hardware in the array.
\end{itemize}

Table \ref{tab:DigitalBackendFactsheet} summarizes the key design parameters of the CHIME Pathfinder's digital backend. In addition to those, the system had to be designed with enough flexibility to allow testing of real-time gain corrections, RFI removal, high-speed
and triggered data tapping for ancillary science such as pulsar and radio
transient signal analysis, and beamforming along each cylinder. The design is also required to be scalable to 2560 inputs
for the full CHIME instrument.

The hardware, firmware and software components of the digital backend are described in more detail in the
sections below.

\begin{table}[htb]
    \begin{center}
        \begin{tabular}{l|l}
        \hline
        Number of analog inputs & 256 \\
        \hline
        Analog sampling & 800\,MSps @ 8 bits \\
        \hline
        Channelizer Type & 2048 sample PFB/FFT\\
        Frequency channels & 1024 bins, 390\,kHz/bin \\
        Channelizer data path 
            & Input: 8 bits \\
            & Internal: 18+18 bits complex \\
            & Output: 4+4 bits complex \\
        \hline
        Power Consumption & Channelizer: 1.2\,kW\\
                          & Correlator:  10\,kW \\
        \hline
        Data rates &  Digitized analog inputs: 1.64 Tbps    \\
            & Shuffle: 1.54 Tbps (Rx+Tx, plus overhead) \\  
            & Output to GPU correlators: 819.2 Gbps (plus overhead) \\ 
            & Output from GPU correlator:  $\sim$100 Mbps (30 second integration) \\
        \hline
        Baselines & 32,896  \\
        \hline
        Computations & 
            Channelizer: $\sim$0.6T complex MAC/s \\
            & Correlator: 13T complex MAC/s \\
        \hline
        \end{tabular}
    \end{center}
    \caption{Key parameters for the Pathfinder digital backend.}
    \label{tab:DigitalBackendFactsheet}
\end{table}

\subsection{Analog-to-digital converter daughterboards}

\begin{figure}[htbp]
    \begin{center}
        \includegraphics[width=\textwidth]{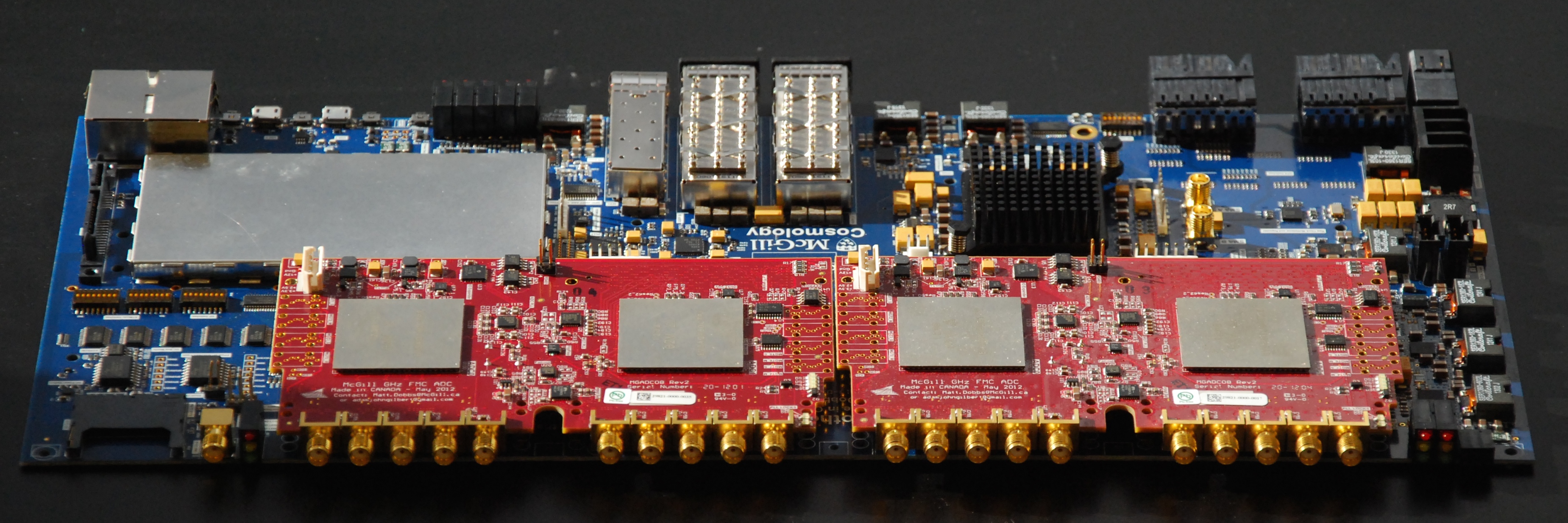}
    \end{center}
    \caption{Two 8-channel, 8-bit,\,1250 MSps, FMC-compatible analog-to-digital daughterboards (red boards) seated on a ICE motherboard (blue board). }
    \label{fig:iceboard}
\end{figure}

The analog-to-digital conversion  of the feed signals is performed using
custom double-wide FPGA mezzanine card (FMC) compliant daughter boards equipped with two E2V
EV8AQ160 analog-to-digital (ADC) chips (see Figure~\ref{fig:iceboard}).  Each
ADC chip has four inputs that can sample at up to 1.25\,GSps at 8\,bits.

The sky signal in the absence of man-made signals is well
encoded with only a few bits, with 4 bits having the effect of increasing any properly amplified white noise
system temperature by $\sim$2 percent.  However, man-made RF power consists of both
broadband bursts and strong narrowband signals which require
additional dynamic range.  Initial testing found that 8\,bit sampling will
provide the dynamic range to sample the sky adequately given the RF conditions
at the site.

For CHIME, the sampling rate is set to 800\,MSps, and the 400-800\,MHz sky signal is
directly sampled using the second Nyquist zone. The analog inputs achieve 
more than 15\,dB return-loss from 300\,MHz to 1.1\,GHz.  The input passband is broader, from
150\,MHz to 1.1\,GHz.  These analog components are constrained to one
section of each board, and are behind a shield that forms a Faraday cage.
The analog traces are routed predominantly on internal board layers
to limit cross-talk.  The neighboring inputs have been measured to have less than -50\,dB cross-talk
levels, limited by the ADC chip.

Each daughterboard requires a 10\,MHz clock input either through a SMA connector on the
front panel, or as an LVDS or LVPECL signal from the host motherboard.  The
board has a software-programmable phase locked loop (PLL) which creates the
1.6\,GHz clock that drives both ADCs. A post-PLL ADC clock with less than
500\,fs jitter is needed to ensure that the ADC performance is degraded by less
than 0.1\,bits.

A synchronization input from the motherboard or an SMA connector on
the front panel allows the ADC data acquisition to be started in a precise way
relative to the 10\,MHz clock and ADC clock, which allows acquisition of data frames
with a deterministic phase across the array.

An on-board I$^2$C EEPROM stores the digital serial number information along with individual board
testing history, performance, and parameters.  Also, an SPI bus allows the host
FGPA to read out the temperatures of the ADC cores and the board, as well as
communicate with the ADC control registers.

\subsection{FPGA (``ICE'') Motherboard } 
\label{sec:ice}

The FPGA motherboard (also known as ``ICE motherboard'', shown in
Figure~\ref{fig:iceboard}) accepts two FMC-compliant high-pin-count  CHIME ADC
daughterboards that connect to a single Xilinx Kintex\,7 XC7K420T FPGA. The
choice of the FPGA was driven by requirements in terms of throughput
needed for data shuffling and GPU links, Input/Output (I/O) pin count to the
FMCs, backplane and other ICE motherboard subsystems, logic resources (gates, RAM,
DSP blocks),  cost, and power consumption.

The FPGA board has 19 high-speed (12.5\,Gbps GTX) serial links
interfaced through the backplane
connector to support the Pathfinder and full CHIME data shuffling. It also
has two QSFP+ connectors which connect to a GPU node and one SFP+ connector which connects to a
control computer. The FPGA can also receive trigger, synchronization and 
GPS-based timestamp signals from the backplane.

The board offers an identifying EEPROM, temperature sensors, FMC power
control, and voltage/current monitoring for every power rail.  This allows for 
self discovery and diagnostics of a large set of motherboards.

Each ICE motherboard is equipped with a Texas Instruments AM3871 ARM processor with 1
gigabyte of DDR3 DRAM, an SD card slot, two gigabit Ethernet ports, USB, UART, and SATA
connections.  The processor runs a Linux operating system, allowing for 
remote-programming of the FPGA as well as  providing always-on monitoring of the
hardware and an arbitrated network-based control interface to the FPGA
subsystems.

The clocking for the entire board is derived from one 10\,MHz clock which is
received from the backplane. An on-board crystal oscillator and a front panel SMA connector are
also available for single-board or bench top development work. The clocking is
then distributed through a low-jitter network to the FMC mezzanine boards and to
the two onboard PLLs which create multiple clocks to drive the FPGA and ARM
processor.

The ICE motherboards are designed for a 9U standard VME physical crate design. They
are 14 inches by 6.5 inches, and can be spaced by 0.8 inch with the
ADC mezzanines installed.  The FPGAs and ADCs require active cooling provided
by the host crate system.  Each board requires approximately 75\,W of power
when running the Pathfinder firmware with two ADC mezzanines. The board
operates from a single supply in the range of 14-20\,V, and the nominal 2\,MHz
buck converters can be synchronized by the FPGA to restrict the switching noise
to known frequencies.

\subsection{FPGA Firmware}

The firmware that operates on the FPGA is mostly custom code written in VHDL
and simulated and compiled using the latest Xilinx Vivado software suite. This
approach enabled us to maximize the use of the Kintex 7 FPGA and ICE motherboard potential as the
new tools could easily place, route and meet timing closure in
half the time required for older Xilinx tools on large FPGA designs.

The firmware is subdivided into modules that interface with each other using the
industry-standard AXI4-Streaming bus protocol to carry the data and control
signals. 

All firmware configurations can be set talking directly to the FPGA using a simple UDP packet structure, or by communicating through the ARM processor.  

The FPGA has no embedded processor and any high-level
functions and algorithms are performed by custom control software on either an external control computer or the ARM processor.

In addition to the command system, the core firmware provides all the
resources needed to operate the ICE motherboard independently of the ARM processor.
This includes access to the buck regulators, I$^2$C-based devices, the FMC
hardware, etc. An internal frequency counter monitors internal and external
clocks to confirm proper operation of the signal processing chain. 

The first signal processing module performs the data acquisition from the FMC
ADC boards. The data is acquired at 800\,MSps through 8 LVDS lines and a 400
MHz DDR clock. The module aligns the data acquisition of each line with a 78\,ps resolution
to compensate for the board and FPGA routing delays. The data is deserialized
and combined into a 200\,MHz, 32-bit wide  AXI4 stream that is passed on to the
channelizer module as frames of 2048 8-bit samples. The data acquisition
module provides logic to 
deterministically start data framing on a known edge of the ADC sample clock relative to the 10 MHz reference.

The channelizer module starts its signal processing by selecting its source stream from the ADC or from an
integrated test pattern generator. The stream is fed to a customized poly-phase filter bank
(PFB) and fast-Fourier transform (FFT) that has been generated by the
CASPER\footnote{https://casper.berkeley.edu/} toolset and has been wrapped in
an AXI interface.  The PFB includes a sinc-Hann window applied to 4 data
frames, and outputs a frame of 1024 complex frequency samples in a 18+18 bit
format. The following scaler module applies a 16+16 bit complex gain to each
frequency bin. The complex gain is stored in two tables that can be configured
and switched in real time in order to account for system gain and delay variations. The
result is finally scaled to (4+4) bit complex values and saturates instead of folding. The output data is accompanied with ADC, FFT and scaling saturation
flags which can be used to
identify the strongest broadband and line radio frequency interference. A
FPGA-based statistics subsystem also keeps track of ADC and scaling
saturations for independent data monitoring.

The data streams from the 16 channelizers are internally reordered by a
crossbar module that aligns the incoming data streams and selects and routes
specific frequency bins from every input to one of its 16 output streams.
Each output stream is typically configured to contain a subset of 64
frequency bins from all channelizers. The channel selection map is fully
configurable to allow exclusion of unusable frequencies (due to RFI) and to
adjust the downstream bandwidth. The crossbar repacks the data and saturation
flags into larger blocks to increase the efficiency of data transfers on the
GPU host.

The data shuffling module takes 15 of the 16 reordered streams and sends
them to every other board in the crate over the backplane using 10\,Gbps
links implemented using the FPGA's high-speed serial GTX transceivers. The data is encoded in
64B/66B format, is scrambled to balance the DC content of the data, and is
encapsulated in simple packets with a cyclic redundancy check (CRC) code to
detect transmission errors. The data coming from the corresponding boards is
also received in the same way. After the full transaction, one FPGA now
possesses a subset of 64 frequency bins from all 256 channelizers of the crate.

The 16 data streams are then passed through another crossbar that rearranges
the data into 8 output streams, each containing 8 frequency bins from
all the channelizers of the array. An array of 8 custom 10 gigabit Ethernet
UDP packet transmitters is then used to send these 8 data streams to a GPU
node through the two QSFP+ connectors. Each packet is
accompanied with a header identifying the data source, the format and size
of the payload data, and a timestamp.

\subsection{Crate and backplane} 
\label{sec:backplane} 

\begin{figure}[htbp]
    \begin{center}
        \subfloat[]{\includegraphics[height=1.7in]{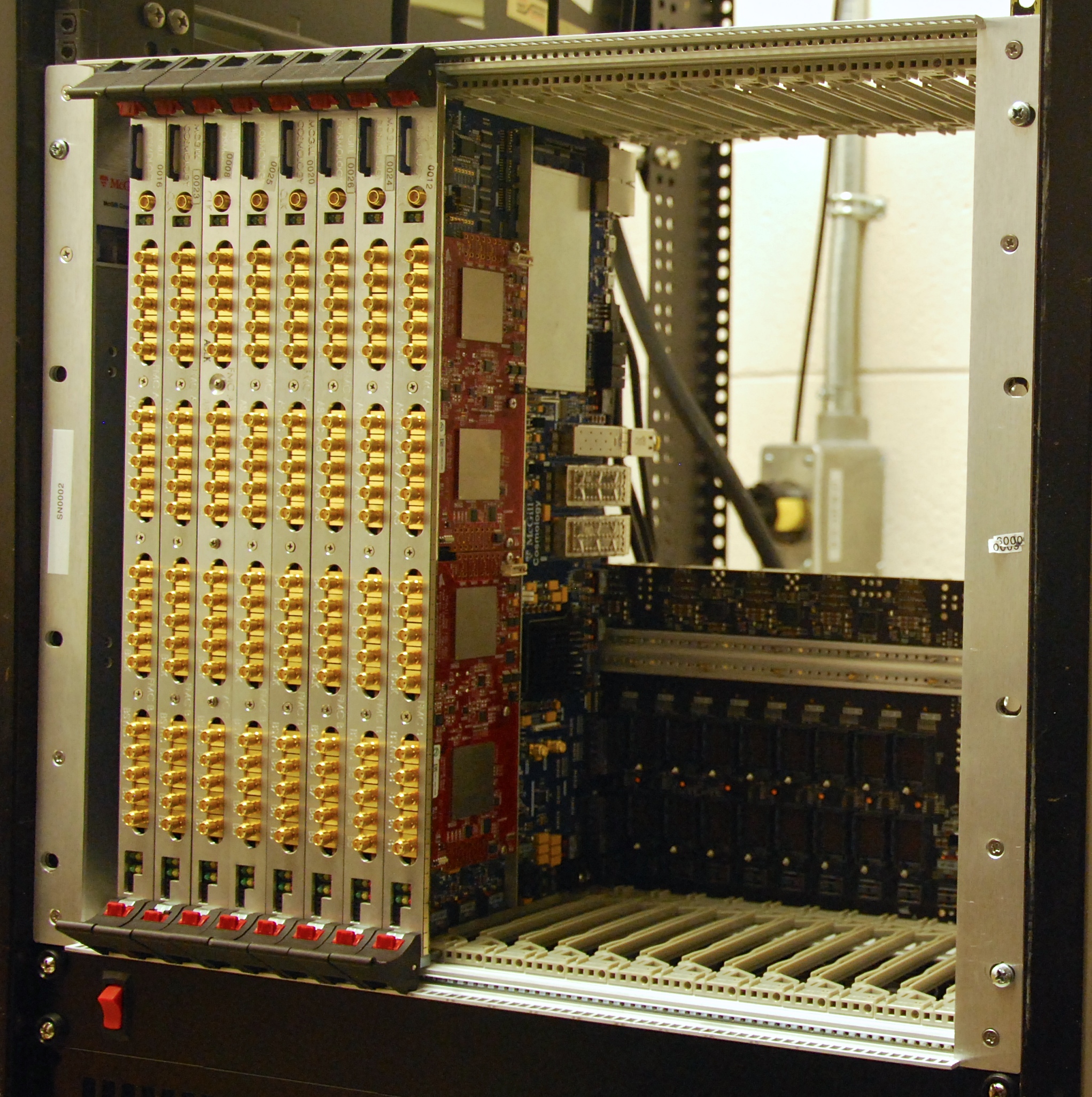}}
        \subfloat[]{\includegraphics[height=1.7in]{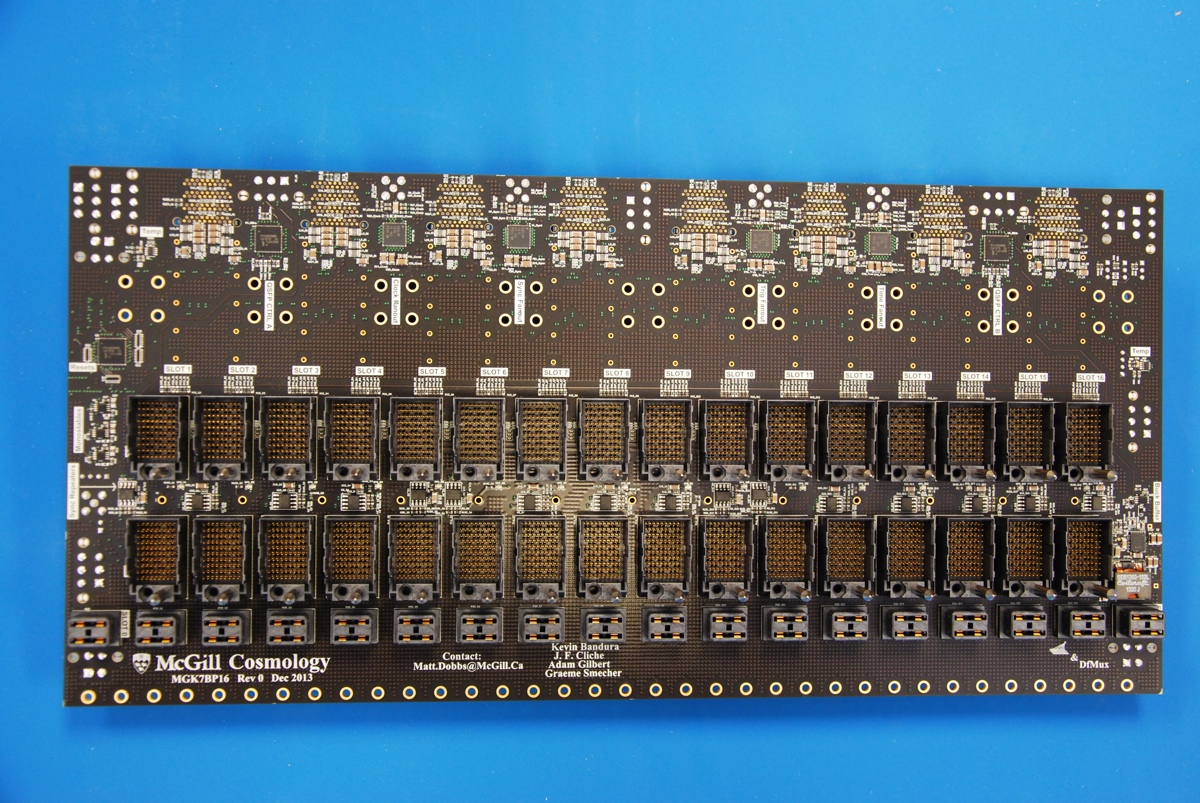}}
        \subfloat[]{\includegraphics[height=1.7in]{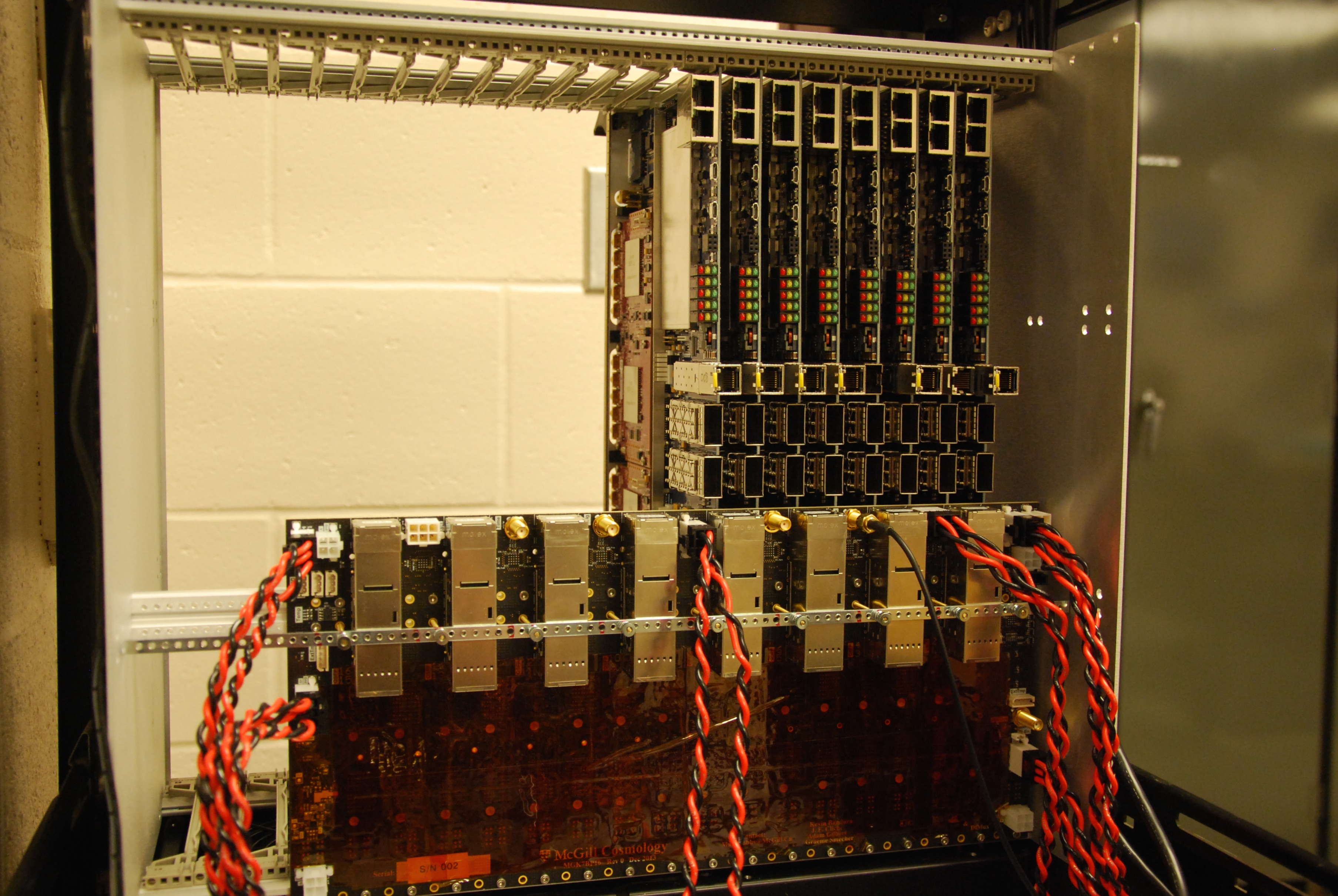}}
    \end{center}
    \caption{(a) 9U crate partially populated with 8 ICE motherboards. (b) Front image of the 
    custom 16-slot full-mesh high-speed backplane.  (c) Image of the rear of crate,
    showing the backplane's power entry cables, 16 QSFP+ connections and SMA
    inputs for clock, trigger, timestamp and synchronization distribution to
    the motherboards. The image also shows the rear-accessible ports provided
    by each ICE motherboard: 2x Gigabit Ethernet, reset buttons, USB/UART, SFP+ and
    2x QSFP+.}
    \label{fig:backplane}
\end{figure}

The CHIME Pathfinder channelizer consists of a single crate containing
16 ICE motherboards. The crate is 9U high and uses standard VME mechanics but a custom high-speed backplane, shown in Figure~\ref{fig:backplane}.

The backplane provides a low-jitter clock distribution system to the
motherboards from a single SMA CMOS level input.  It also distributes three
additional digital signals: the ADC synchronization pulse, trigger pulse, and a
timestamp signal. An I$^2$C interface allows any motherboard or an external
controller to remotely power down or reset any board individually, and allows
the backplane temperature and power to be monitored.

A key feature of the backplane is its 10\,Gbps full-mesh networking that
connects every board with every other one in both directions. This mesh
connects directly to the motherboard's FPGA multi-gigabit transceivers to
implement the low-cost passive shuffling network. Extreme care was taken in
maximizing the signal integrity of the 10\,Gbps links. To achieve this,
25\,Gbps Molex Impact connectors (Model 761657107) are used to mate with
the motherboards, and transmission line discontinuities are reduced by using single-plane direct connections and
back-drilled vias.  The board is made with
Panasonic Megtron 6, a low-loss material compatible with more standard board
assembly, and very low profile (VLP) copper foil is used to further reduce losses. The
board is laid out to reduce cross-talk between signals in the network. Strong
and weak signaling levels are kept well apart and are shielded by intermediate
power planes. The FPGA's internal dynamic equalizer compensates for the
low-pass frequency response and remaining distortions caused by the links. Preliminary
tests between the two furthest boards (slots 1 and 16) show
error-free transmission at 10\,Gbps. 

The backplane offers 16 QSFP+ connectors connected directly to the motherboard
FPGAs to provide an additional 640 Gbps of off-crate data transfer through
copper or optical cables, and will be used for the full CHIME data
shuffling between 5 crates. The QSFP+ can be interrogated to confirm cable
connectivity and LEDs can be controlled to assist manual wiring and diagnosis.

\subsection{GPU correlator}

Visibility calculation and time averaging of all baselines takes place in a
dedicated GPU-based computing cluster.  A diagram of the system is shown in 
Figure~\ref{fig:gpuBlock}.  

The operation is split across 16 fully independent and identical processing
nodes, each responsible for processing the full set of baselines for 1/16th
(25\,MHz) of the CHIME bandwidth. A single control system is housed in the
same cluster, which serves the software and operating system used by the
diskless nodes.  This same system aggregates and buffers the data prior to long-term archiving on a data server.  

Each node is housed in a 4U rackmount chassis and built primarily of high-end
consumer-level components.  The processing takes place in two AMD r9 280x GPUs
and one r9 270x GPU. A pair of enterprise network interface cards (NICs)
receives a total of eight 10 gigabit network connections, streaming a total of
51.2\,Gbps of radiometric data, along with associated headers and flags. This
data rate sets the requirements for most system components.

The incoming data is transferred over a third generation PCIe bus (8 lanes for
each of the network boards) into primary system memory. Headers and flags are
stripped from the data for further processing.  Each packet header has a
sequence number which allows the system to track and manage packet loss, and a
stream ID identifies the frequencies in the packet.   The sequence number is
identical across all links for a given ADC sampling period, so it is used to
provide timing and synchronization between hosts.   The flags are used to track
and correct for ADC and scalar overflows in the data samples.

The data is then DMA transferred from system memory into GPU buffers in large (256MB) blocks.  Each GPU hosts
3GB of on-board RAM (2\,GB for the 270x), used to buffer incoming data prior to
processing. For maximum computational flexibility, 3 connections are distributed
to each r9 280x, 2 connections to the r9 270x -- this results in all boards
operating at roughly 2/3 utilization while performing the $N^2$ correlation. The
transfer is controlled by an Intel i7-4820k CPU, allowing 40 total lanes of
PCIe-3 communications. An EVGA x79 Dark motherboard was chosen to allow 8 lanes
to each of the 5 expansion boards (2xNIC, 2x r9 280x, 1x r9 270x). A primary
bottleneck in the system was found to be the CPU-memory interconnect, and DDR3
2133\,MHz overclocked RAM is used to maximize performance.

The correlation operation takes place in a custom processing kernel written in the OpenCL language\footnote{https://www.khronos.org/opencl/}. Details of this
kernel will be presented in a future paper; we describe it briefly here. A single
instance of the kernel (an OpenCL ``Work Item,'' WI) computes 4x4 correlations
and accumulates them over 256 time steps, roughly 0.6ms. These WIs are grouped
into sets of 64 (OpenCL ``Work Groups'') which share high-speed local memory and
compute a 32x32 correlation block. The 256x256 matrix of correlations is divided
into these 32x32 sub-blocks, and the 36 upper-triangle blocks are computed and
accumulated. (The remaining 28 blocks contain no additional information, due to
the symmetry of correlations.) Efficient operation requires that all
calculations be pipelined as multiply-accumulate operations (MACs), and the
algorithm is able to operate efficiently by using integer operations and packing
two 4-bit values into each register. This packing sets the 256-timestep
accumulation period, and requires a handful of book-keeping operations to take
place at the end of each MAC loop,  accounting for the 8 least significant bit
(LSB) offset on each sample, and accumulating the real and complex portions into
32-bit buffers. Multiple kernel invocations result in longer accumulations, with
correlation buffers read out for archiving at 10-30s cadence.

The control system collects the output correlations over a gigabit Ethernet
network, merges the data and archives it locally to an array of 3TB hard drives.
This array can buffer at most several days of data, with the long-term archive
stored in another local building, on a much larger array of disks.  
Custom-written software registers  basic information about data products in a 
MySQL database for indexing  purposes. The same software also manages automatic
transfer of data  products between acquisition computers and long-term
storage and analysis nodes.

Cooling of the GPUs is a significant concern. Consumer-level GPUs
are assembled and sold by a variety of vendors (using the same basic layout and
identical processors), with various solutions for heat dissipation. A variety of
brands were tested for thermal performance, and Sapphire ``Dual-X'' branded boards were
chosen for both the r9 280x and r9 270x, due to their remarkable cooling
capabilities (keeping the processor die to roughly 70\,C under load, 20-30\,C cooler than all
other models tested in our setup). We are presently exploring direct-to-chip
watercooling options, and anticipate retrofitting the system in the coming
months.

\section{Observation Strategies and Challenges}

The CHIME Pathfinder observing strategy is quite simple, since the instrument
cannot move.  Each day the telescope observes the whole northern sky, with the
observation time for every source set by the east-west field of view of
2.5$^\circ$-1.3$^\circ$ and the source declination: $\mathrm{obstime} /
\mathrm{day} = 4\,\mathrm{min} \times  \mathrm{fov} \times \cos(\mathrm{dec})$.
The full correlation matrix is saved at a 30 second cadence.

One of the challenges of the system comes from the calibration needed in order to detect the BAO with the CHIME Pathfinder. From previous simulations of a Pathfinder-like
instrument\cite{Shaw:2014vy} all telescope primary beams must be
known to 0.1\% and the system gain for each feed must be known to 1\% in
order to be able to properly reconstruct the power spectrum.  
See the accompanying SPIE proceedings\cite{newburgh2014} for full details
about the calibration of CHIME, which will be implemented on the
Pathfinder. A brief  summary of the calibration procedures planned to achieve these levels of instrument stability is given below.

\begin{itemize}

\item Measure the primary beam using bright point sources and pulsars.  The
bright points sources will give a first measure of an overall complex gain
calibration and beam-width as a function of frequency.  Pulsars with their
inherent on-off period allow one to remove all signals from that data that do
not pulse at the frequency of the pulsar.  The beam will be measured further
using pulsar holography by additionally correlating the signal from the DRAO 26m
Telescope tracking the pulsars.

\item Inject a broadband calibration signal.  We will  inject
into every feed a broadband calibration signal, and measure and correlate that
signal with all the CHIME feeds.  This low-level injected signal is switched on
and off on a $\sim$ second timescale. The switched signal is then used to measure
and adjust for the complex relative gain of every feed, removing complex gain
changes on scales longer than the switching time.

\item Use the redundant baseline information.  The CHIME Pathfinder is a very
redundant interferometer by design.  These redundant baselines can be used to
calculate the complex gain of each feed without having any previous knowledge
of the sky.

\end{itemize}

Another remaining challenge lies in interference management. Even though the
CHIME Pathfinder is located in the radio astronomy reserve of the Dominion
Radio Astrophysical Observatory, there are still man-made radio frequency
signals which will interfere with observing the sky.  The CHIME Pathfinder
plans to use real-time RFI flagging and excision techniques to mitigate those.
The most extreme events which saturate the digital signal are flagged by the
FPGA system and are passed to the GPU X-engine.  Further processing will be
performed on an intermediate integration scale of milliseconds
to further process the data looking for excessive excursions of the signal.
Post-processing will handle the final data clean-up.

\section{Conclusion}

\begin{figure}[tb]
    \begin{center}
        \includegraphics[width = 0.5\textwidth]{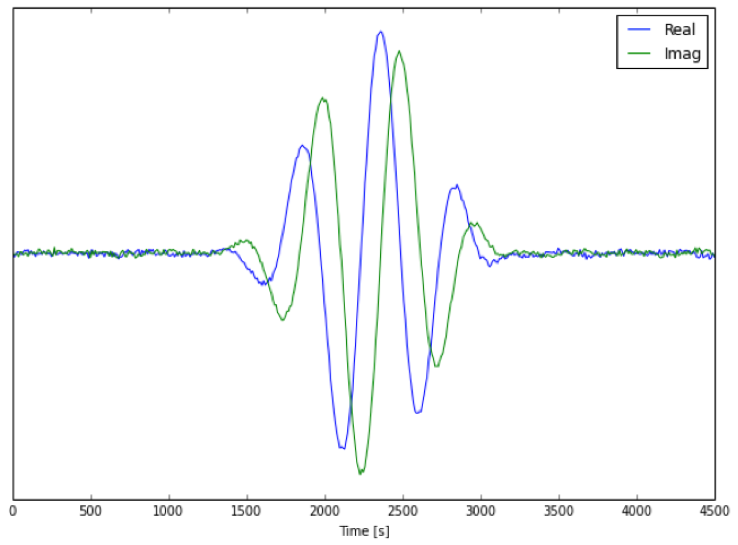}
    \end{center}
    \caption{First fringes from an east-west CHIME Pathfinder baseline of .  Shown 
    are the real and imaginary parts of a single visibility for a single 
    frequency channel.  }
    \label{fig:firstLight}
\end{figure}

The CHIME Pathfinder is currently being commissioned at the Dominion
Radio Astrophysical Observatory.
The cylinder structures
have been completed and the analog and digital electronics are being phased in
with 16 feed channels currently installed. The first fringes observed
with the telescope, from Cassiopia A, are shown in Figure~\ref{fig:firstLight}.  We are comparing measured sky maps to our galactic model to assess system temperature.  We have performed holography with the DRAO 26m telescope to measure our beam shape.  Finally, we have installed a broadband signal injection system to monitor complex gain in real-time.

The CHIME Pathfinder is paving the way for all the aspects of the CHIME
experiment.  It is a 1/7th area version of the full system with 1/10th of the
analog and digital electronics. All components being tested are to be scaled
up to the full system.  The Pathfinder will also be used to explore beyond the
baseline design by investigating RF-over-fiber links \cite{2013JInst...810003M}, and real-time
FFT beam-forming with the FPGAs and GPUs. Meanwhile, ground is currently being broken in preparation
for the full CHIME construction.

\appendix    

\acknowledgments     

We are very grateful for the warm reception and skillful
help we have received from the staff of the Dominion Radio
Astrophysical Observatory, operated by the National Research Council Canada.

We acknowledge support from the Canada Foundation for Innovation, 
the Natural Sciences and Engineering
Research Council of Canada, the
B.C. Knowledge Development Fund,  le Cofinancement gouvernement du
Qu\'ebec-FCI, the Ontario Research Fund, the CIfAR Cosmology and
Gravity program, the Canada Research Chairs program, and the National
Research Council of Canada. 
M.\ Deng acknowledges a MITACS
fellowship.  
 We thank Xilinx  and the XUP for their generous donations.  


\bibliography{bibliography}   
\bibliographystyle{spiebib}   

\end{document}